\documentclass[twocolumn,prd,aps,groupedaddress,nopacs,nofootinbib]{revtex4-1}

\usepackage{amsmath}
\usepackage{amssymb}
\usepackage{latexsym}
\usepackage{graphicx}
\usepackage{hyperref}
\usepackage{mathrsfs}
\usepackage{color}
\usepackage{bm}
\usepackage[caption=false]{subfig}

\begin{document}
 
\title{Null tests of the cosmological constant using supernovae}
 \author{Sahba Yahya$^1$, Marina Seikel$^{1}$, Chris Clarkson$^2$, Roy Maartens$^{1,3}$, Mathew Smith$^1$ 
}
\affiliation{\it $^1$Physics Department,
University of the Western Cape,
Cape Town 7535,
South Africa
\\
\it $^2$Department of Mathematics \& Applied Mathematics, University of Cape Town, Cape Town 7701, South Africa,\\
 \it $^3$Institute of Cosmology \& Gravitation, University of Portsmouth, Portsmouth PO1 3FX, UK.
 }

\begin{abstract}

The standard concordance model of the Universe is based on the
cosmological constant as the driver of accelerating expansion. This
concordance model is being subjected to a growing range of
inter-locking observations. In addition to using generic observational
tests, one can also design tests that target the specific properties
of the cosmological constant.  These null tests do not rely on
parametrizations of observables, but focus on quantities that are
constant only if dark energy is a cosmological constant.  We use
supernova data in null tests that are based on the luminosity
distance.  In order to extract derivatives of the distance in a
model-independent way, we use Gaussian Processes.  We find that the
concordance model is compatible with the Union 2.1 data, but the error
bars are fairly large. Simulated datasets are generated for the DES
supernova survey and we show that this survey will allow for a sharper
null test of the cosmological constant if we assume the Universe is
flat. Allowing for spatial curvature degrades the power of the null
test.
    
\end{abstract}

\maketitle

\section{Introduction}

The simplest model that can explain the apparent acceleration of the
Universe is the `concordance' $\Lambda$CDM model, with $\Omega_{{m}}
\approx 0.3$ and zero spatial curvature $\Omega_{{K}}= 0$.  The
concordance model is consistent with all observations to date~\cite{Ade:2013zuv}. Current
observations favour a dark energy model with equation of state $w(z)
\approx -1$, although there are modified gravity models with no dark energy that are
also consistent with the data \cite{Clifton:2011jh}. Next-generation
experiments such as DES \cite{Abbott:2005bi}, LSST
\cite{Abell:2009aa}, EUCLID \cite{Laureijs:2011gra} and the SKA
\cite{Blake:2004pb}, are expected to dramatically improve on current
constraints and introduce new observables.

It is typical to parametrize $w(z)$ in order to differentiate between
various dark energy models, or to parametrize background and
perturbation variables to test classes of modified gravity models.  A
complementary approach is to test the consistency of the concordance
model itself, independent of the values of $\Omega_{{m}}$ and
$\Omega_{{K}}$. A range of null tests designed specifically to probe
various aspects of the concordance model have been introduced (see
e.g. \cite{Clarkson:2007pz, Uzan:2008qp, Sahni:2008xx, Zunckel:2008ti,
  Shafieloo:2009hi, Clarkson:2009jq, Nesseris:2010ep, Seikel:2012cs}
and \cite{Clarkson:2012bg} for a review).
 
Type Ia Supernovae (SNIa) are the best distance indicators to probe
the expansion history of the Universe. These `standardizable candles' can
be observed to high redshift, and have produced convincing evidence
that the Universe has undergone a recent phase of accelerated
expansion.  Current samples of SNIa (e.g. \cite{Conley:2011ku,
  Suzuki:2011hu, Campbell:2012hi, Scolnic:2013aya, Rest:2013bya}) comprise several hundred SNIa with
$z<1.8$. Forthcoming surveys of SNIa, such as DES
\cite{Bernstein:2011zf}, will produce well-measured light-curves for
over 4000 SNIa, improving the cosmological constraints by an order of
magnitude.

In this paper we use luminosity distances $d_L(z)$ determined from
SNIa observations to test the consistency of the concordance model,
through a set of null tests.  Reconstructing the expansion history of
the Universe in a model-independent fashion is essential for these
tests. To do this, we use Gaussian Processes (GP), which have
previously been used to reconstruct $w(z)$ from SNIa luminosity
distances \cite{Holsclaw:2010nb, Holsclaw:2010sk, Holsclaw:2011wi,
  Seikel:2012uu, Shafieloo:2012ht}. Our analysis is built on
\cite{Seikel:2012cs}, which used $H(z)$ data from the baryon acoustic oscillation (BAO) scale and galaxy ages
to
test the validity of the concordance model.  We use GaPP (Gaussian
Processes in
Python)\footnote{\href{http://www.acgc.uct.ac.za/~seikel/GAPP/index.html}{\url{http://www.acgc.uct.ac.za/~seikel/GAPP/index.html}}},
a package developed by Seikel and introduced in
\cite{Seikel:2012uu}.

The tests based on $H(z)$
are potentially stronger discriminators of the concordance model than
those using SNIa data, since the null tests using $d_L(z)$ require
higher derivative terms than those using $H(z)$. However, null tests
based on direct distance measurements currently have the advantage
that the data sets are much larger and the errors are smaller.

\section{Null tests of $\Lambda$CDM -- theory}\label{theory}

The Friedmann equation, 
\begin{eqnarray} 
&&\frac{H^2(z)}{H^2_0}=\Omega_{{m}}(1+z)^3 + \Omega_{{K}}(1+z)^2 \nonumber\\
&&~~+ (1-\Omega_{{m}} - \Omega_{{K}})\exp\left[3 \int_0^z \frac{1+w(z{'})}{1+z{'}} dz{'}\right]\!,
\label{hz}
\end{eqnarray}
determines the Hubble rate $H$ in terms of today's values for the
density parameters for matter $\Omega_{{m}}$ and curvature
$\Omega_{{K}}$. This is integrated over to obtain the luminosity
distances of SNIa:
\begin{align}
d_L(z) = \frac{(1+z)}{H_0\sqrt{- \Omega_{{K}}}} \sin\left(\sqrt{-\Omega_{{K}}} \int_{0}^{z} \frac{dz'}{H(z')/H_0} \right).
\label{Dz}
\end{align}
The equation of state parameter of dark energy, $w=p_{\rm
  de}/\rho_{\rm de}$, can be expressed in terms of the dimensionless
comoving luminosity distance,
\begin{equation}
D(z) \equiv H_0(1+z)^{-1} d_L(z),
\end{equation}
as \cite{Starobinsky:1998fr,Nakamura:1998mt,Huterer:1998qv}:
\begin{eqnarray}
 &&\!\!\!\! w(z)=\Big\{2(1+z)(1+\Omega_{{K}}D^2)D''-[(1+z)^2\Omega_{{K}}D'^2 
 \nonumber \\&&{}\!\!\!\!
+2(1+z)\Omega_{{K}}DD'-3(1+\Omega_{{K}}D^2)]D'\Big\}/ 
\\ \nonumber &&{}\!\!\!\!
\Big\{3\{(1+z)^2[\Omega_{{K}}+(1+z)\Omega_{{m}}]D'^2-(1+\Omega_{{K}}D^2)\}D'\Big\}. 
\label{w}
\end{eqnarray}

Given an observed distance-redshift relationship $D(z)$, it is
possible to reconstruct the equation of state of dark energy and test
the $\Lambda$CDM model \cite{Seikel:2012uu}. However, a disadvantage
of this method is that it depends on the values of the density
parameters, $\Omega_{{m}}$ and $\Omega_{{K}}$, which must be measured
independently \cite{Seikel:2012uu}.

To avoid this problem and test $\Lambda$CDM using SNIa data, we use
the consistency tests introduced in \cite{Zunckel:2008ti} (see also
\cite{Sahni:2008xx,Shafieloo:2009hi}). Following this approach, we
test the null hypothesis that the expansion of the universe can be
described by a flat or a curved $\Lambda$CDM model. 

The assumptions underlying the consistency tests and the null
hypothesis are: (1)~the universe is homogeneous and isotropic on large
scales; (2)~gravity obeys general relativity; (3)~the universe
contains cold matter (with $w=0$) and dark energy. Photons and
neutrinos can be included ($\Omega_\gamma$, $\Omega_\nu$ are known
independently, from CMB data), but it is reasonable to neglect
radiation at the low redshifts probed by SNIa data.  Detection of a
deviation from the consistency tests would imply a violation of at
least one of these assumptions: (1)~large-scale nonlinear
inhomogeneity or anisotropy; (2)~modified gravity; (3)~dynamical dark
energy ($w\neq -1$), or alternatively, a cosmological constant plus an
unknown additional species with equation of state which deviates from
that of cold matter, curvature or vacuum energy.  Any of these
possibilities imply that the standard $\Lambda$CDM is ruled out. Note
that the tests cannot identify which of these possibilities applies.

For a flat concordance model,
i.e. $w=-1$ and $\Omega_{{K}} = 0$, from \eqref{Dz} we find that
\begin{equation}
\Omega_{{m}} \left[(1+z)^3-1\right]D'^2=1-D'^2.
\end{equation}
If we define 
\begin{align}
\label{omh}
 \mathcal{O}_m^{(1)}(z)&~=\frac{1-D'(z)^2}{[(1+z)^3-1]D'(z)^2} ,
\end{align}
then
\begin{equation} 
\mathrm{flat~\Lambda CDM~implies~}  \mathcal{O}_m^{(1)}(z) = \Omega_{{m}}.
\end{equation}
Thus we obtain a null test of the concordance model: 
\begin{equation}
 \mathcal{O}_m^{(1)}(z) \neq \Omega_{{m}} ~\mathrm{falsifies~flat~\Lambda CDM}.
\end{equation}
Any variation of $\mathcal{O}_m^{(1)}(z)$ with redshift reflects an
inconsistency between the flat $\Lambda$CDM model and observations. 
To detect evolution of $\mathcal{O}_m^{(1)}$ with redshift we can differentiate $\mathcal{O}_m^{(1)}(z)$, from which we define
the additional diagnostic:
\begin{align}
\label{eq:L}
\mathcal{L}^{(1)}(z)&~=(1+z)^{-6} \Big\{2\left[(1+z)^3 -1 \right] D''(z) \nonumber\\
            &~~+3(1+z)^2D'(z) \left[1-D'(z)^2\right]\Big\},
\end{align}
which vanishes if and only if $d\mathcal{O}_m^{(1)}/dz=0$. The factor $(1+z)^{-6}$ (which was not used in \cite{Zunckel:2008ti}),
ensures stability of the errors (see below).
If $\mathcal{L}^{(1)}$ is nonzero at any redshift, then observations
are incompatible with $\Lambda$CDM:
\begin{equation}
\mathcal{L}^{(1)}\neq0~\mathrm{falsifies~flat~\Lambda CDM}.
\end{equation}

We can extend this approach to include spatial curvature, and derive
null tests for general (curved) $\Lambda$CDM. Using \eqref{hz},
\eqref{Dz} and \eqref{w} with $w(z) = -1$, and solving for
$\Omega_{{m}}$ and $\Omega_{{K}}$, we find~\cite{Clarkson:2009jq,
  Clarkson:2012bg}:  
\begin{align}
\label{omwhole}
\Omega_{{m}} &~= 2\Upsilon(z)\Big\{ \left[(1+z)^{2} -D^{2}-1 \right]D''
 \nonumber \\ 
&~~ - \left( D'^2 -1 \right)  \left[( 1+z)D' -D \right] \Big\}\equiv \mathcal{O}_m^{(2)}(z),\\
\label{OKwhole}
\Omega_{{K}} &~= \Upsilon(z)\Big\{ 2\left[ 1-(1+z) ^{3}\right]D''
\nonumber\\ 
&~~ +3D'\left(D'^2-1\right)( 1+z)^{2} \Big\}\equiv \mathcal{O}_{{K}}(z).
\end{align}
Here $\Upsilon(z)$ is defined by
\begin{align}
\Upsilon^{-1} &~ = -2\left[ 1-(1+z) ^{3} \right]  D^2 D'' \nonumber\\ 
&~ -\Big\{  (1+z)\left[  ( 1+z)^{3}-3( 1+z)+2 \right]  D'^{2} \nonumber\\
 &~ -2\left[ 1-( 1+z)^{3} \right] DD' -3(1+z)^{2}D^{2} \Big\}D'.
\end{align}
Then we have
\begin{eqnarray}
&& \mathcal{O}_m^{(2)}(z) \neq \Omega_{{m}}~\mathrm{falsifies~curved~\Lambda CDM}, \\
&& \mathcal{O}_{{K}}^{(2)}(z) \neq \Omega_{{K}}~\mathrm{falsifies~curved~\Lambda CDM}.
\end{eqnarray}

These are not independent tests: the derivative of
$\mathcal{O}_{{K}}^{(2)}$ vanishes if and only if the derivative of
$\mathcal{O}_m^{(2)}$ vanishes. Hence we need only a single diagnostic
for vanishing derivative. We use the derivative of
$\mathcal{O}_m^{(2)}$ to define
\begin{align}
\label{L_{{K}}2}
 \mathcal{L}^{(2)} &~ = (1+z)^{-6} D'^2 \Big\{D \Big[-3\left(1+z\right)
 \nonumber\\
 &~ \times \left(D'^2-1\right) \left(2D' + 3 (1+z)D''\right)
 \nonumber\\
 &~ + 2zD'''\Big(3+z(3+z)\Big)\Big] + 9(1+z)^2D^2D''^2 
 \nonumber\\
 &~ + 3(1+z)D^2D' \Big(2D'' -(1+z)D'''\Big)
 \nonumber\\
 &~ +  6(1+z)^2D'^2 \left(D'^2-1 \right)-\Big[3z^2(3+z)D''^2 
 \nonumber\\
&~  + zD'\Big(z(3+z)D''' - 6(2+z)D''\Big)\Big](1+z)\Big\},
\end{align}
which vanishes if and only if $d\mathcal{O}_m^{(2)}/dz=0$. (Again we
use the pre-factor to stabilize the errors.) 
Then we have the null test for curved $\Lambda$CDM:
\begin{equation}
\mathcal{L}^{(2)}(z)\neq 0~\mathrm{falsifies~curved~\Lambda CDM}.
\end{equation}

In principle, $\mathcal{L}^{(1)}$ and $\mathcal{L}^{(2)}$ provide no additional
information compared to $\mathcal{O}_m^{(1)}$ and
$\mathcal{O}_m^{(2)}$. However, it is easier to detect a deviation
from zero than to confirm that a quantity is constant, especially
since the exact value of this constant is not known a priori. The
disadvantage of $\mathcal{L}^{(1)}$ and $\mathcal{L}^{(2)}$ is that
they require higher derivatives than $\mathcal{O}_m^{(1)}$ and
$\mathcal{O}_m^{(2)}$, which are more challenging to constrain.

Another problem with  $\mathcal{L}^{(1)}$ and $\mathcal{L}^{(2)}$ is
the degeneracy between $w$ and $\Omega_m$: a model with redshift
dependent $w$ can be formally consistent with $\Lambda$CDM within the
error bars of the reconstruction if the value of $\Omega_m$ is
adjusted accordingly. Such cases can only be identified with the
$\mathcal{O}_m$ tests, but not with $\mathcal{L}$ (see section
\ref{Discussion} for details).

Note that $\mathcal{L}^{(1)}$ and $\mathcal{L}^{(2)}$ are not
identical to $d\mathcal{O}_m^{(1)}/dz$ and $d\mathcal{O}_m^{(2)}/dz$,
respectively. Starting from these two derivatives, we have neglected
the denominators, which add significant noise to the tests without
adding extra information, and used a pre-factor $(1+z)^{-6}$ to obtain
$\mathcal{L}^{(1)}$ and $\mathcal{L}^{(2)}$. We are free to do this
without loss of generality, since we are testing the equality of these
quantities with zero. As a consequence, the error bands of the
reconstructions do not necessarily increase with redshift as one might
expect, and the size of the errors of $\mathcal{L}^{(1)}$ and
$\mathcal{L}^{(2)}$ are not directly comparable. In addition, the errors added from extra redshift factors are small when we have spectroscopic redshift measurements.

\section{Null tests using SNIa data}\label{reconstruction}

To apply these null tests using current datasets, it is essential to
choose a model-independent method to reconstruct $D(z)$ and its
derivatives. For this purpose, we use GP (via the 
 GaPP code \cite{Seikel:2012uu})
to smooth the data and
reconstruct the derivatives.

\subsection{Gaussian Processes}

GP provide a distribution over functions that are suitable to describe
the data. At each point $z_i$, the distribution of function values
$f(z_i)$ is a Gaussian. Thus the reconstruction consists of a mean
function with Gaussian error bands. The function values at different
points are correlated by a covariance function $k(z,\tilde{z})$, which
depends on a set of hyperparameters (e.g.\ the characteristic length
scale $\ell$ and the signal variance $\sigma_f$). This also provides a
robust way to estimate derivatives of the function in a stable manner.

In contrast to parametric methods, GP do not assume a specific form
for the reconstructed function. Instead only typical changes of the
function are considered. The hyperparameter $\ell$ corresponds roughly
to the distance one needs to move in input space before the function
value changes significantly, while $\sigma_f$ describes typical
changes in the function value. 

The choice of covariance function affects the reconstruction to some
extent. A general purpose covariance function is the squared
exponential covariance function $k(z,\tilde{z}) = \sigma_f^2
\exp\left[-(z -\tilde{z})^2/(2 \ell^2)\right]$. However, 
here we use the Mat\'ern ($\nu=9/2$) covariance function:
\begin{eqnarray}
k(z,\tilde{z}) &=& \sigma_f^2
  \exp\Big(-\frac{3\,|z-\tilde{z}|}{\ell}\Big) \nonumber \\
  &&~\times \Big[1 +
  \frac{3\,|z-\tilde{z}|}{\ell} + \frac{27(z-\tilde{z})^2}{7\ell^2}  \nonumber\\
&&~~~~~~
+ \frac{18\,|z-\tilde{z}|^3}{7\ell^3} +
  \frac{27(z-\tilde{z})^4}{35\ell^4} \Big]. \label{mat}
\end{eqnarray}

For a given covariance function, the probability distribution of the
hyperparameters depends only on the data.  It is necessary either to
marginalize over the hyperparameters $\sigma_f$ and $\ell$, or to fix
the hyperparameters to their maximum likelihood values. Here we choose
the latter approach, which is a good approximation and computationally
much less expensive than marginalization.

We choose the Mat\'ern ($\nu=9/2$) covariance function because it
leads to the most reliable results amongst the covariance functions
that we have tested. Here, ``reliable'' means the following: For
various assumed cosmological models and many realizations of mock data
sets, the assumed model {\em on average} lies within the reconstructed
1-$\sigma$ limits for approximately 68\% of the redshift range (and
within the reconstructed 2-$\sigma$ limits for $\sim95$\% of the
redshift range). These values are theoretically expected, thus making
Mat\'ern ($\nu=9/2$) a reliable covariance function for our
purposes. A detailed analysis regarding the optimal choice of
covariance function can be found in \cite{Seikel:2013fda}. (Note that
these results only apply to GP reconstructions using $D$
measurements. When applying GP to other data, another covariance
function might be more reliable.)

We follow \cite{Seikel:2012uu,Seikel:2012cs}, which contain a
summary of the technical details of GP. The only difference in our
approach here is that we use the Mat\'ern covariance function
\eqref{mat} instead of the squared exponential. (For detailed reviews
of GP, see \cite{Rasmussen, MacKay}.)

\subsection{Application to real data}\label{real data}

We now apply GP to the Union 2.1 dataset \cite{Suzuki:2011hu} and
determine the current constraints on the consistency of
$\Lambda$CDM. This data set comprises $580$ SNIa, with $0.015<z<1.5$,
and includes a covariance matrix which incorporates a systematic
uncertainty. 

The distance modulus, $\mu=m-M$, is the difference between the observed magnitude~$m(z)$ and the absolute magnitude of an object~$M$, and is given by
\begin{equation}
 \mu(z) + 5 \log H_0 - 25 =  5 \log \left[(1+z)D(z)\right].
 \label{MtoD}
\end{equation}
We choose $H_0= 70\,$kms$^{-1}$Mpc$^{-1}$, as in~
\cite{Suzuki:2011hu}. Note that  $H_0$ and $M$ are degenerate in \eqref{MtoD} so we can
fix $H_0$ and only consider the uncertainties in $M$ which are
included in the covariance matrix of the Union 2.1 dataset~\cite{Suzuki:2011hu}~-- this includes the errors on $H_0$. We convert $\mu$ to $D$ and add the
theoretical values $D(z=0)=0$ and $D'(z=0)=1$ to the data set.

\begin{figure*}
\includegraphics[width=0.4\textwidth]{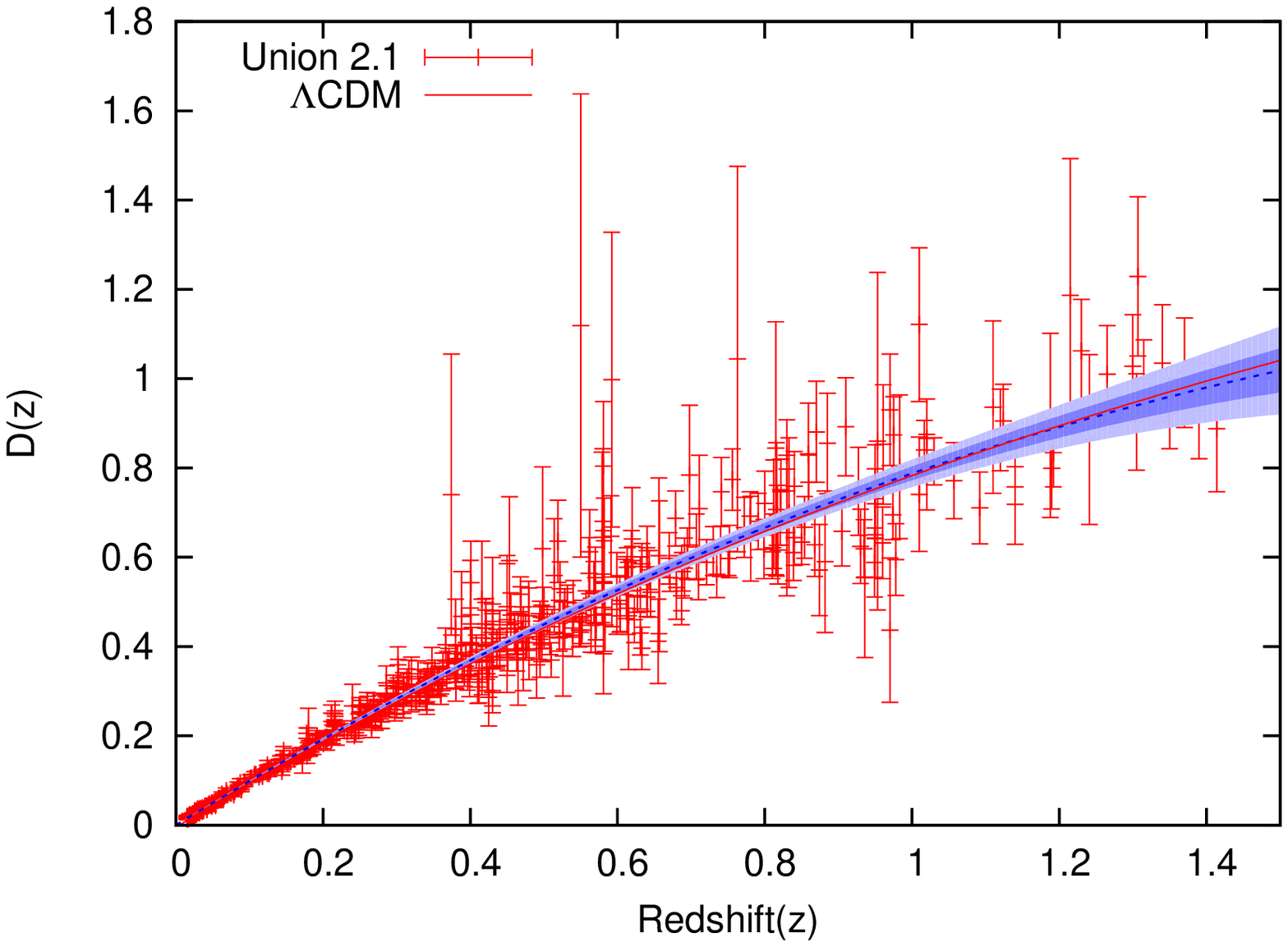}\quad
\includegraphics[width=0.4\textwidth]{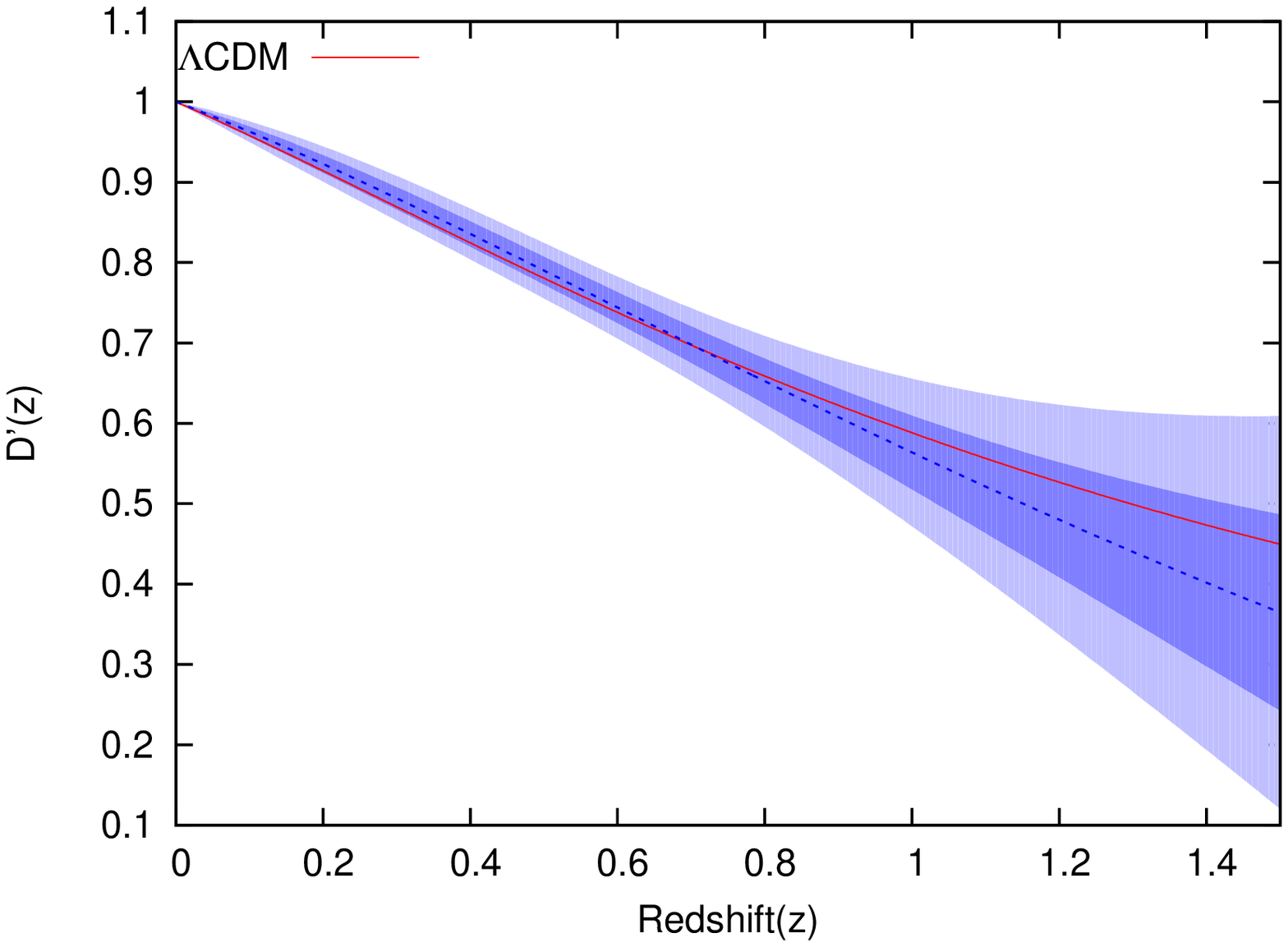}\\
\includegraphics[width=0.4\textwidth]{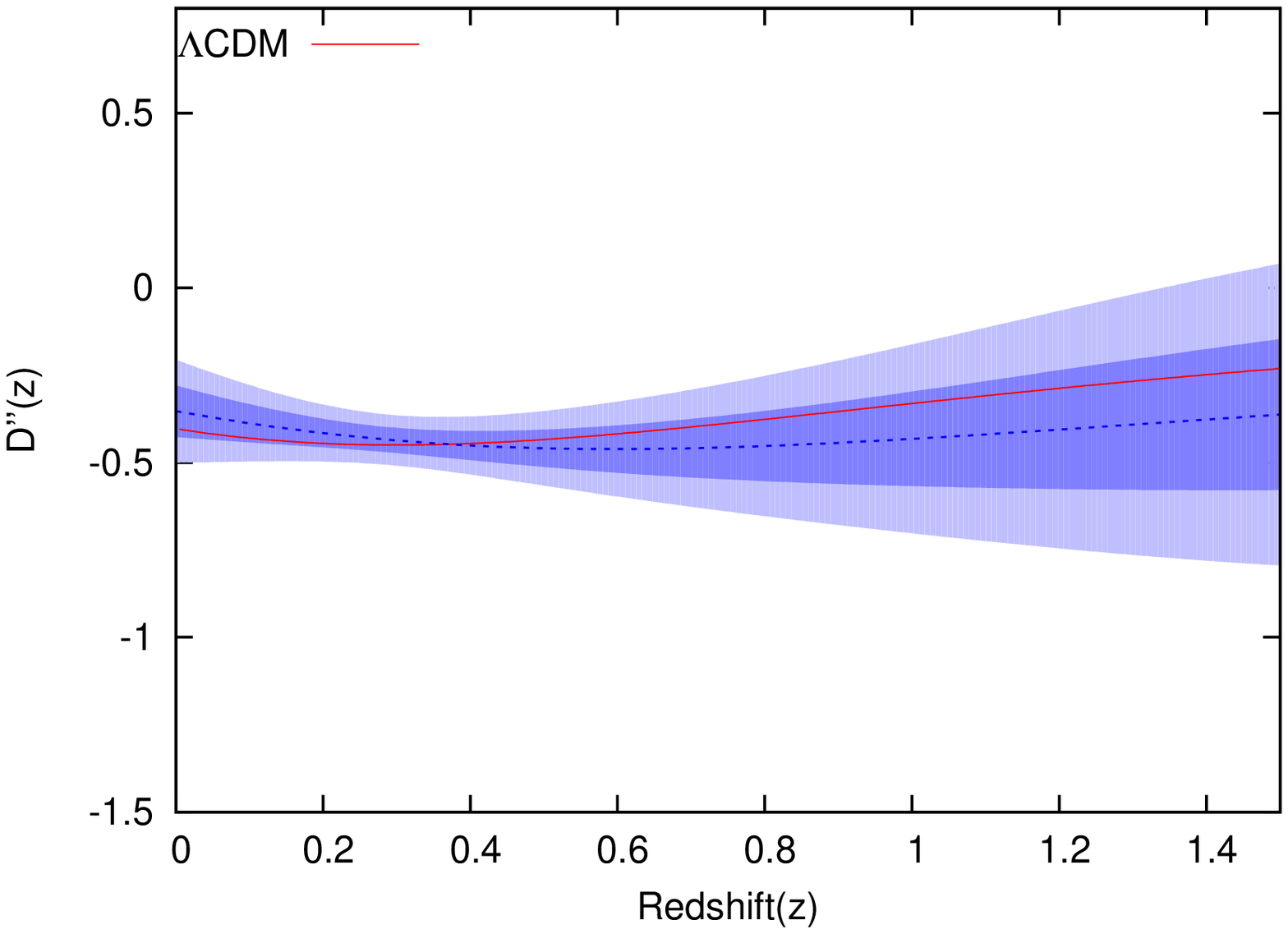}\quad
\includegraphics[width=0.4\textwidth]{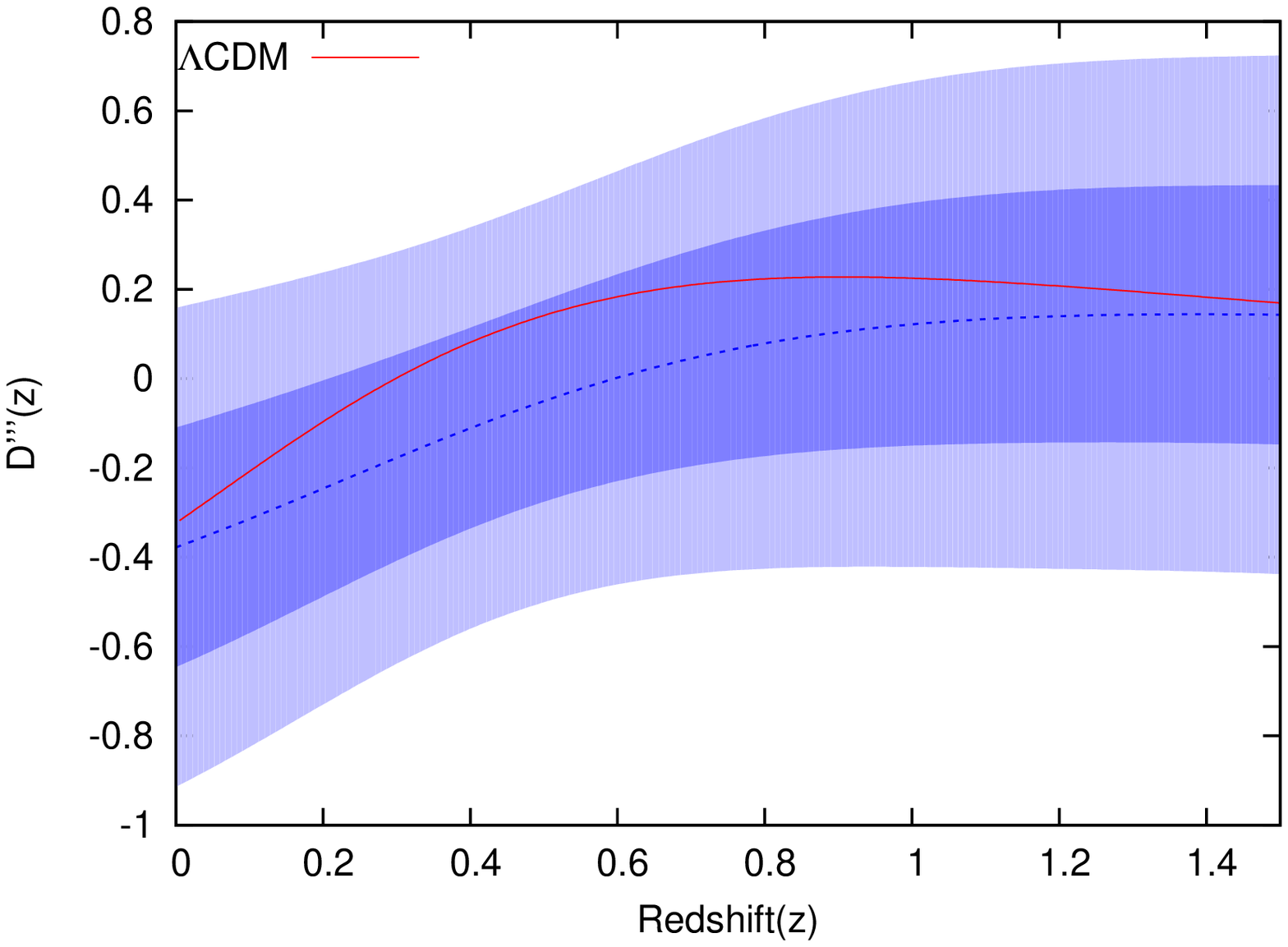}
\caption{Gaussian processes reconstruction of $D$, $D'$ ({\em top})
  and $D''$, $D'''$ ({\em bottom}) for Union 2.1 data.  The red
  (solid) line is flat $\Lambda$CDM with $\Omega_{{m}}=0.27$. The blue
  (dashed) line is the mean of the reconstruction. Shaded areas give
  $95\%$ (light) and $68\%$ (dark) confidence limits of the
  reconstructed function.}
\label{Fig:salt2_mlcs17}
\end{figure*}

\begin{figure*}
\includegraphics[width=0.3\textwidth]{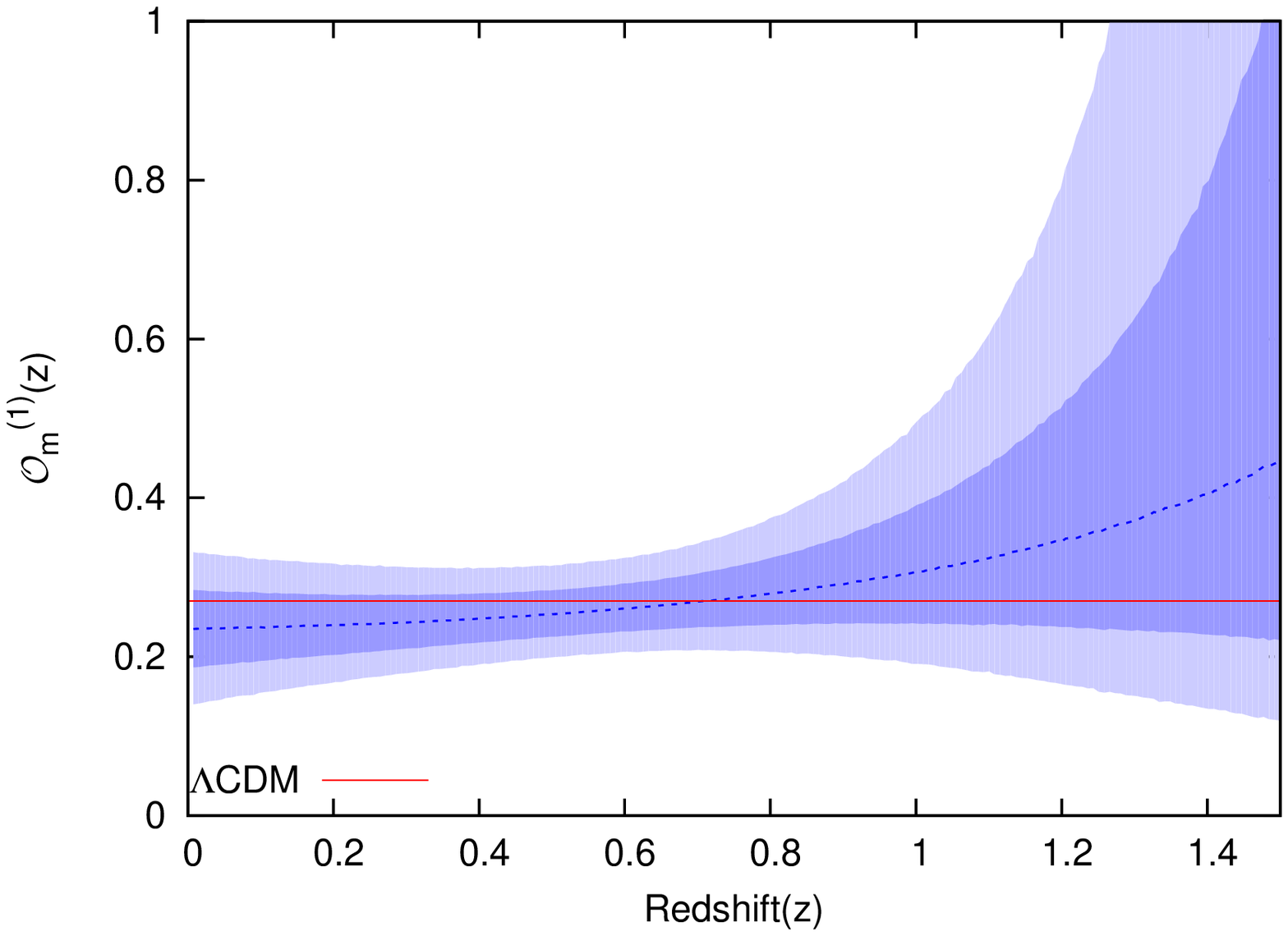}\quad
\includegraphics[width=0.3\textwidth]{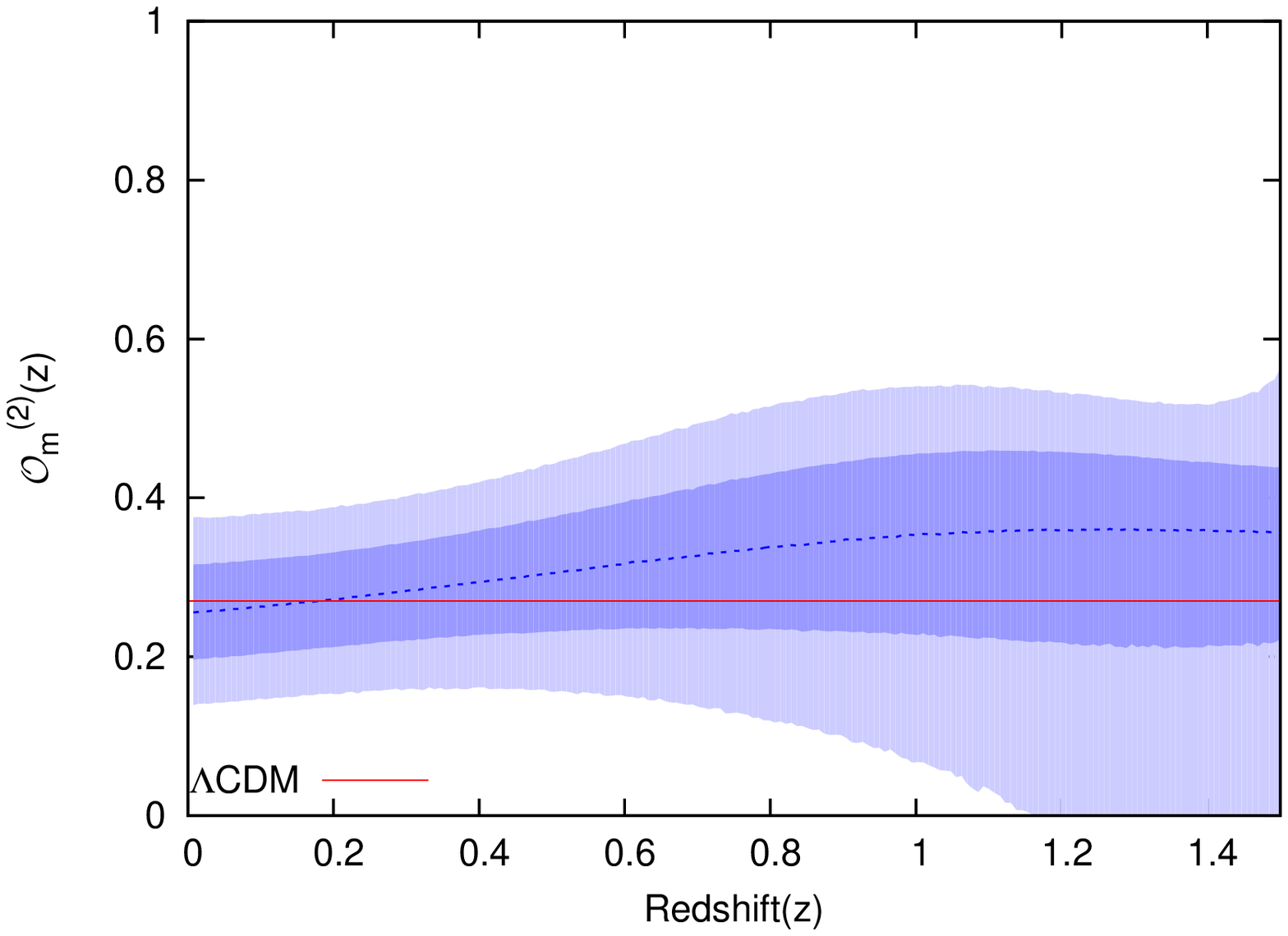}\quad
\includegraphics[width=0.3\textwidth]{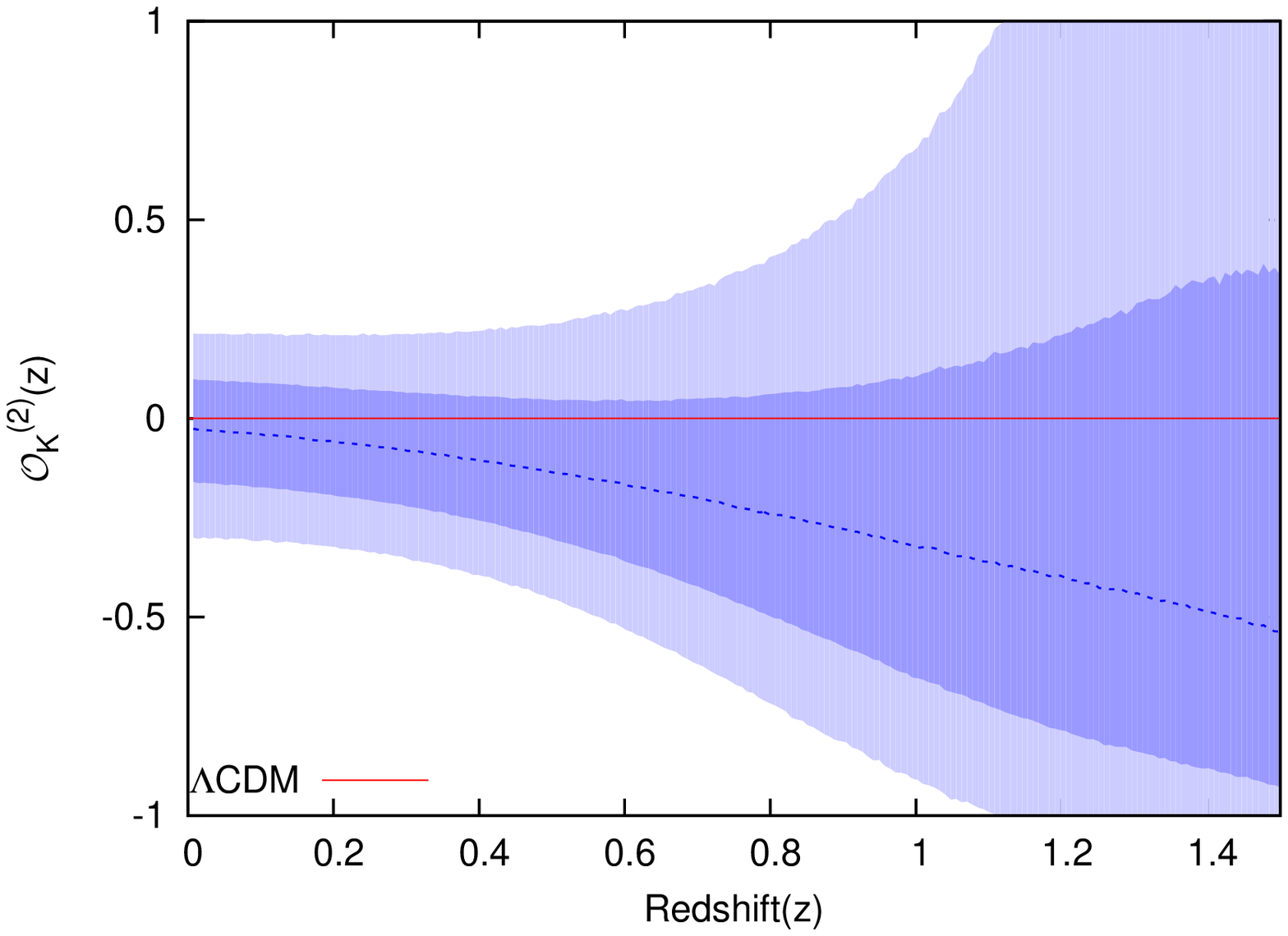}
\caption{Reconstruction of $\mathcal{O}_m^{(1)}$ ({\em left}),
  $\mathcal{O}_m^{(2)}$ ({\em middle}) and $\mathcal{O}_K^{(2)}$ ({\em
    right}) for Union 2.1 data. Lines and shadings are as in Fig. \ref{Fig:salt2_mlcs17}.
}
\label{Fig:om1}
\end{figure*}

\begin{figure*}
\includegraphics[width=0.4\textwidth]{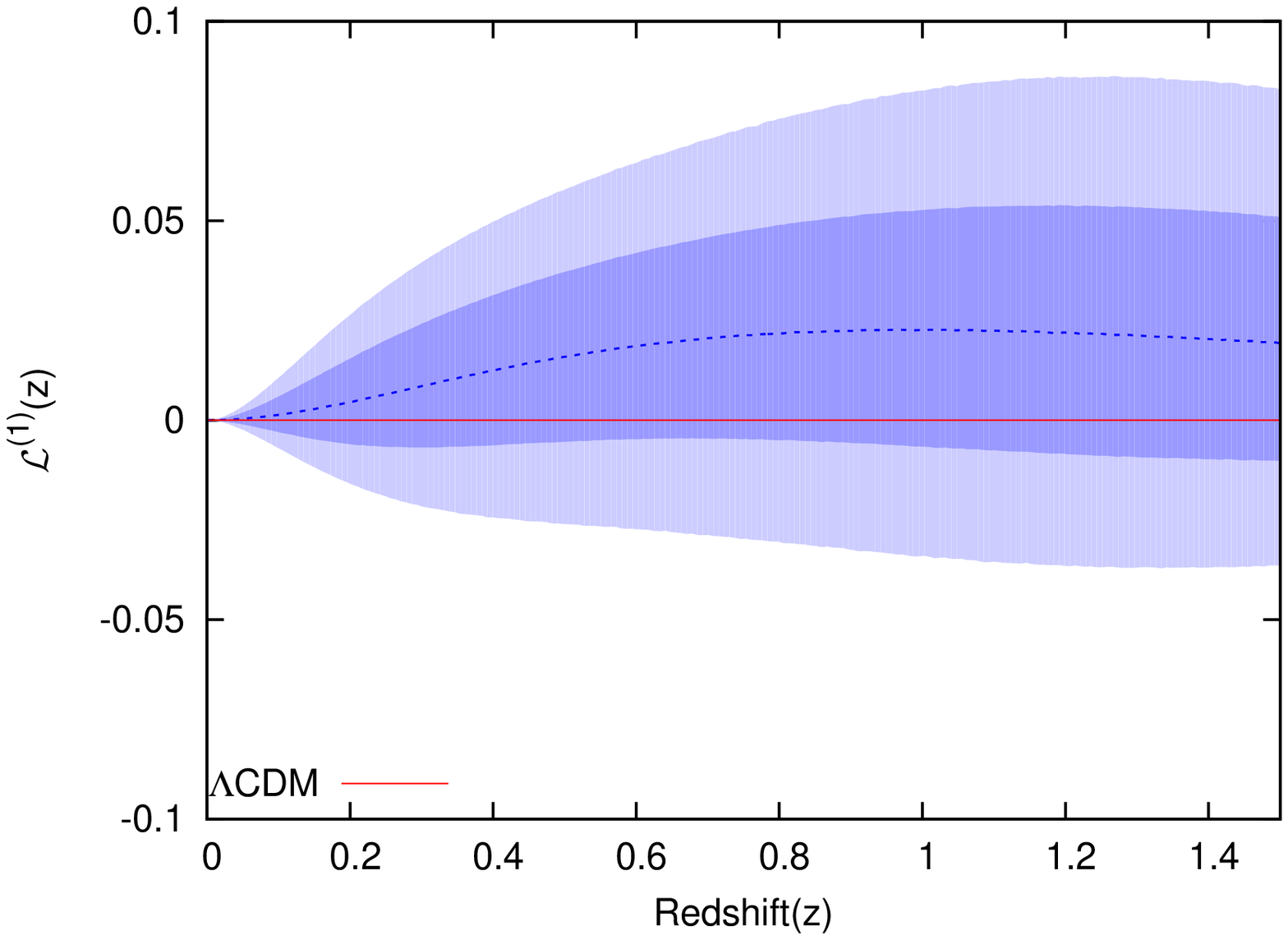}\quad
\includegraphics[width=0.4\textwidth]{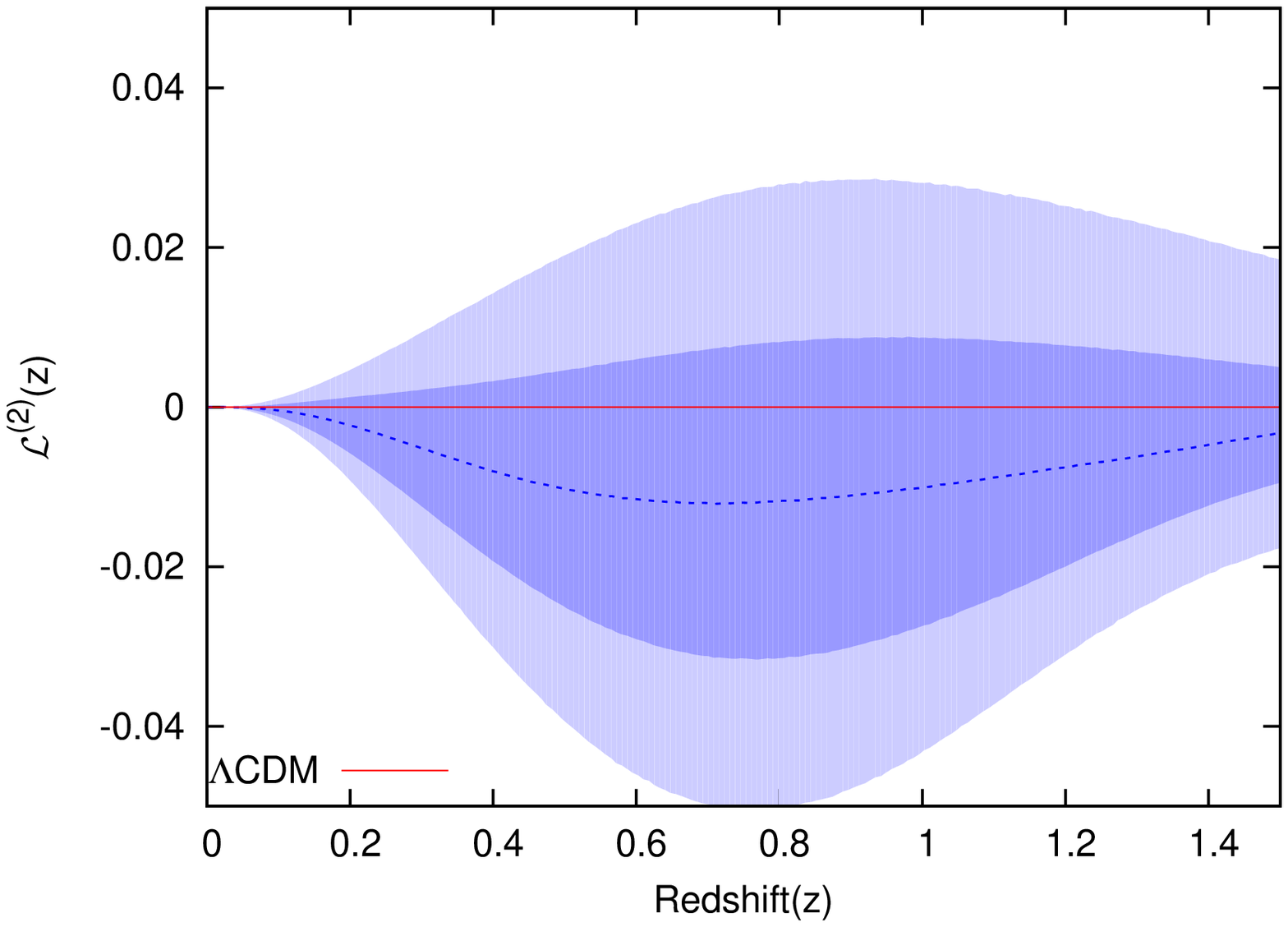}
\caption{Reconstruction of $\mathcal{L}^{(1)}$ ({\em left}) and
  $\mathcal{L}^{(2)}$ ({\em right}) for Union 2.1 data.  Lines and
  shadings are as in Fig. \ref{Fig:salt2_mlcs17}.  }
\label{Fig:lm1}
\end{figure*}

Figure~\ref{Fig:salt2_mlcs17} shows the reconstructed $D(z)$ and its
first three derivatives for the Union 2.1 data set, while
Figure~\ref{Fig:om1} shows the inferred reconstructions for
$\mathcal{O}_m^{(1)}$, $\mathcal{O}_m^{(2)}$ and $\mathcal{O}_K^{(2)}$. 
Figure~\ref{Fig:lm1} shows the reconstruction of $\mathcal{L}^{(1)}$ and $\mathcal{L}^{(2)}$. 

The errors on the reconstructed distances in
Figure~\ref{Fig:salt2_mlcs17} increase with increasing order of
derivative. For example, at $z=1.5$, the standard deviation is 0.05
for the reconstruction of $D$, 0.12 for $D'$, 0.22 for $D''$, and 0.29
for $D'''$. The near-constancy of the errors on $D'''$ reflect the fact
that we are unable to constrain rapid variations (carried via higher
derivatives) on scales below a typical length scale, which is roughly
associated with $\ell$. By using Gaussian Processes the scale $\ell$
and the resulting smoothness of the reconstruction is driven purely by
the data. Where there is insufficient \emph{evidence} for rapid
variations, a smooth function will result, which we see in the second
and third derivatives. Further analysis of the redshift-dependence of
the errors can be found in the appendix.

\subsection{Mock data}\label{Mockdata}

To demonstrate the ability of the null tests to distinguish between
different cosmological models when applied to future SNIa datasets, we
produce mock catalogues for two fiducial models:
\begin{itemize}
\item Flat $\Lambda$CDM
\item Dynamical dark energy model with $\Omega_{{K}} = 0$ and 
  \begin{equation}\label{evolvingw}
    w(z) = \frac{1}{2} \Big\{-1 +  \tanh\Big[3\Big(z-\frac{1}{2}\Big)\Big]\Big\}. 
  \end{equation}
\end{itemize}
We take $\Omega_{{m}} = 0.3$. Using the redshift distribution and
scatter anticipated by the Dark Energy Survey (DES)
\cite{Bernstein:2011zf}, we simulate $\sim 4000$ data points in the
redshift range $0 < z <1.2$. Note that the scatter only includes
statistical errors.

\begin{figure*}
\includegraphics[width=0.3\textwidth]{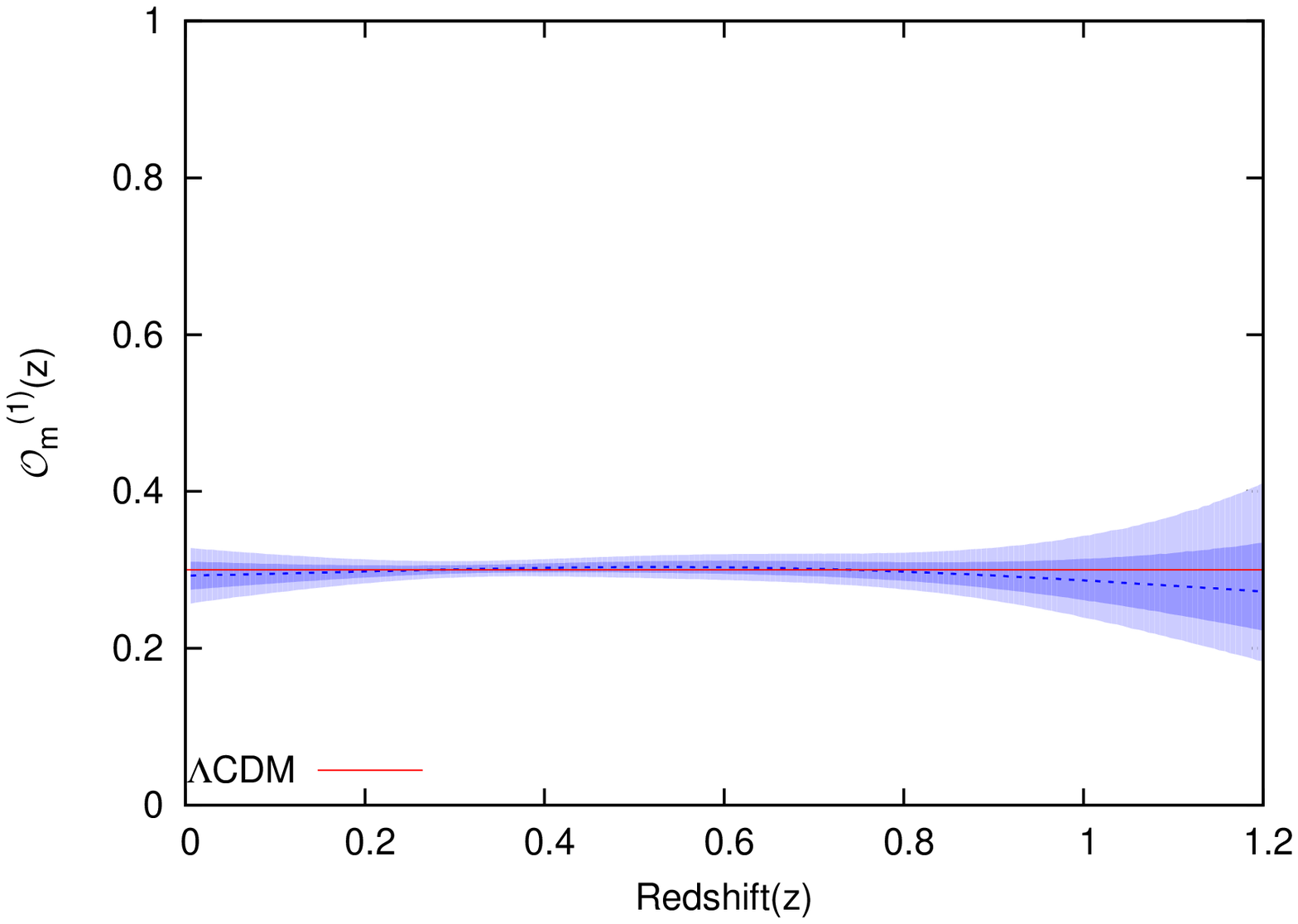}\quad
\includegraphics[width=0.3\textwidth]{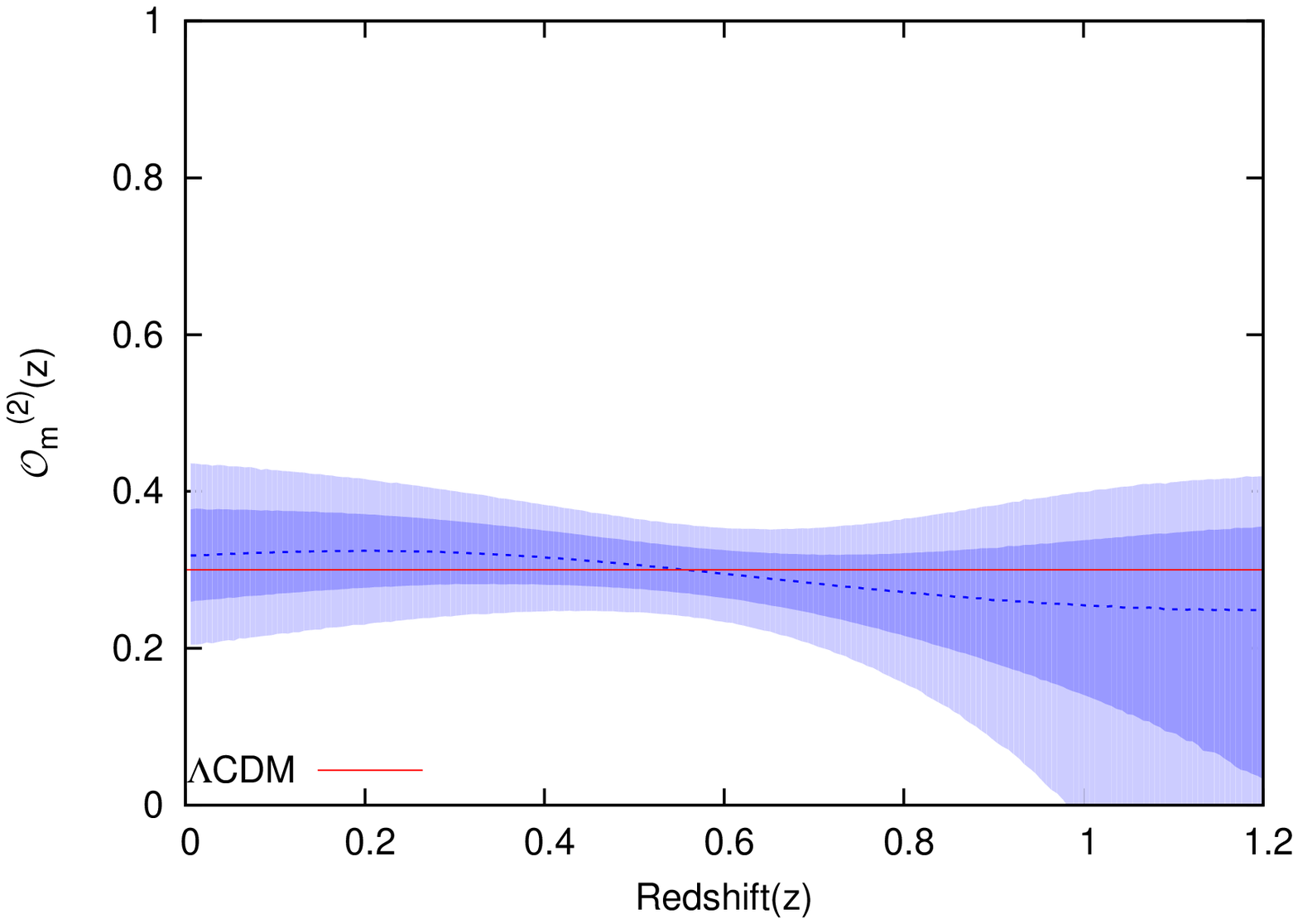}\quad
\includegraphics[width=0.3\textwidth]{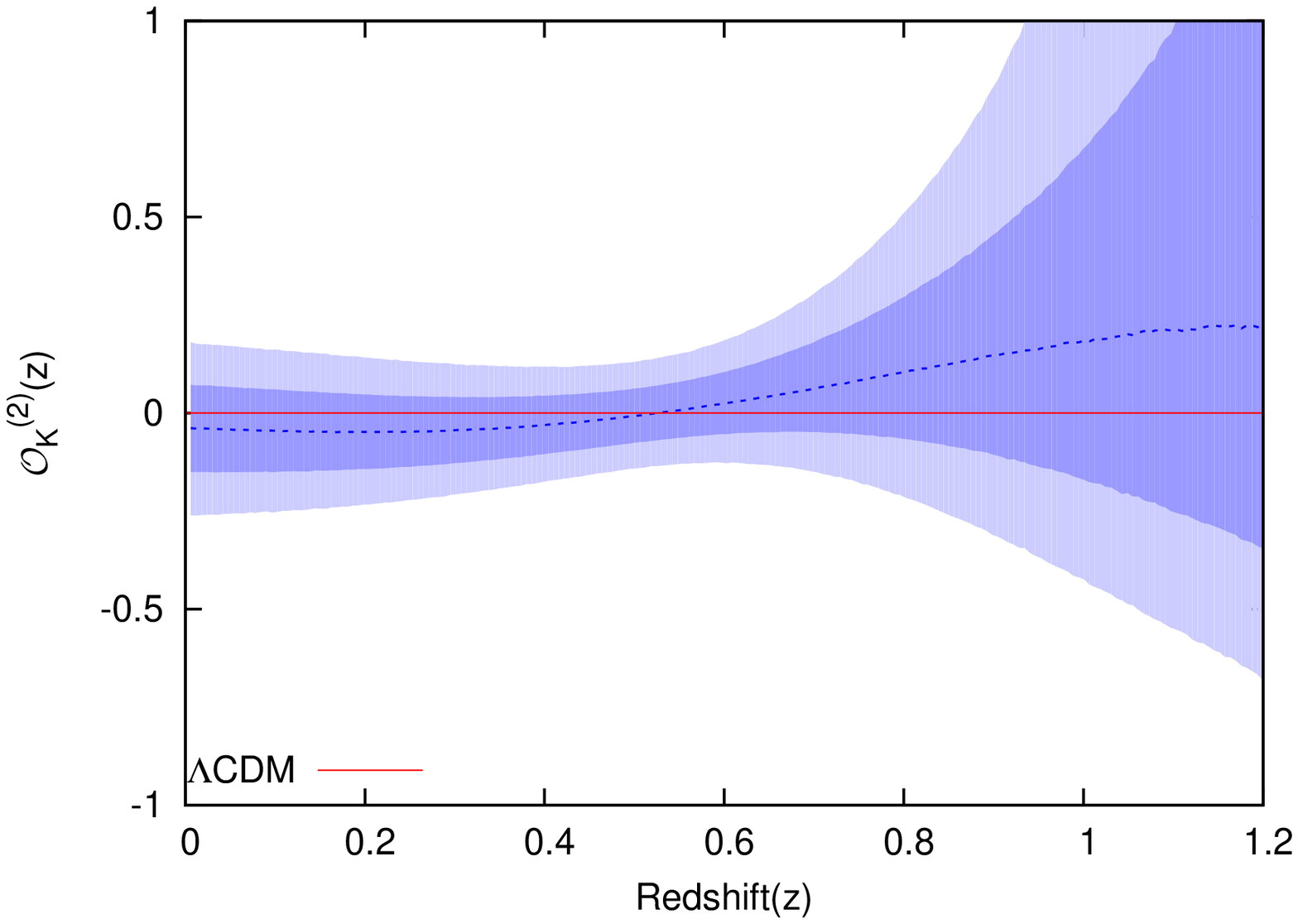}\\
\includegraphics[width=0.3\textwidth]{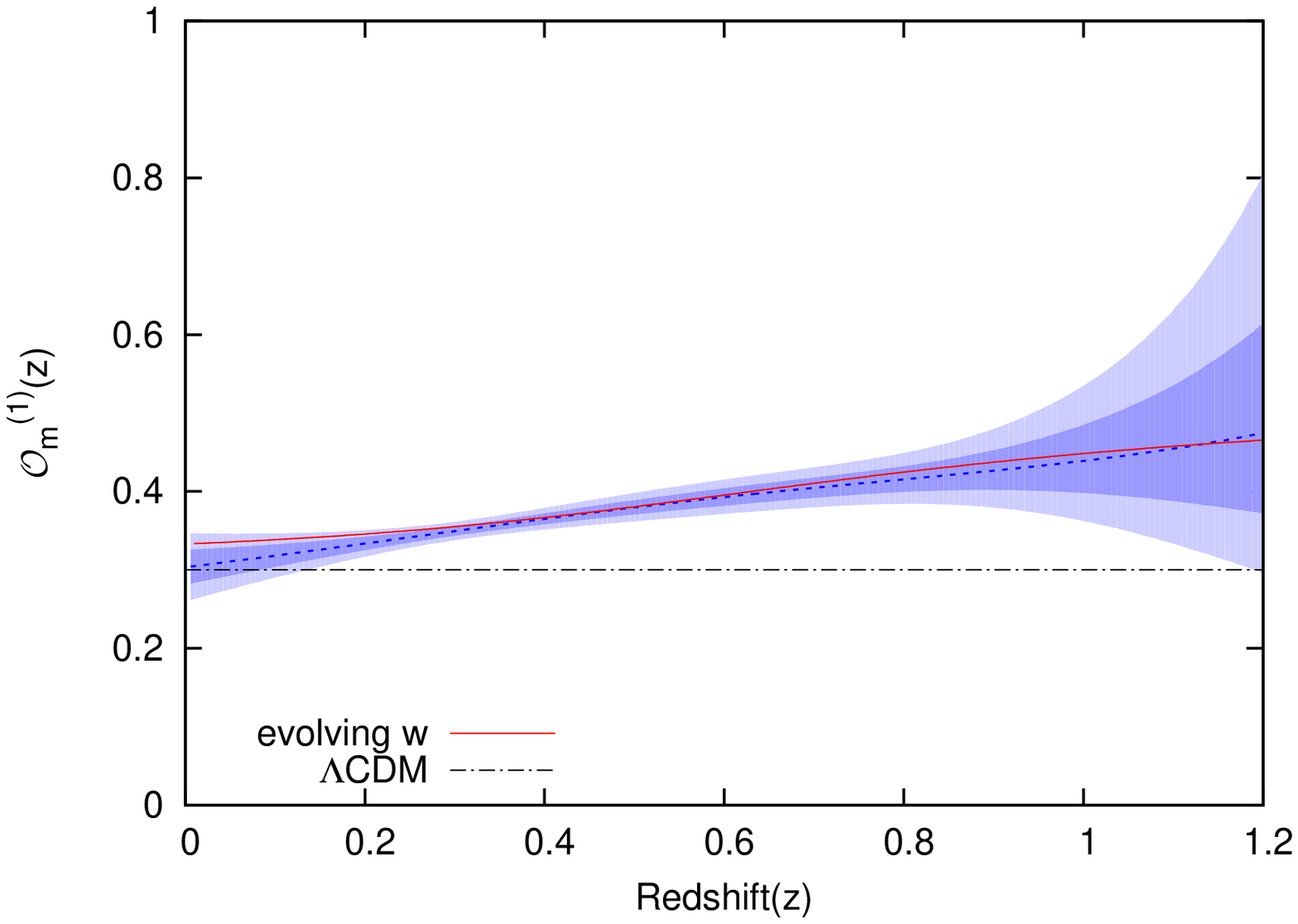}\quad
\includegraphics[width=0.3\textwidth]{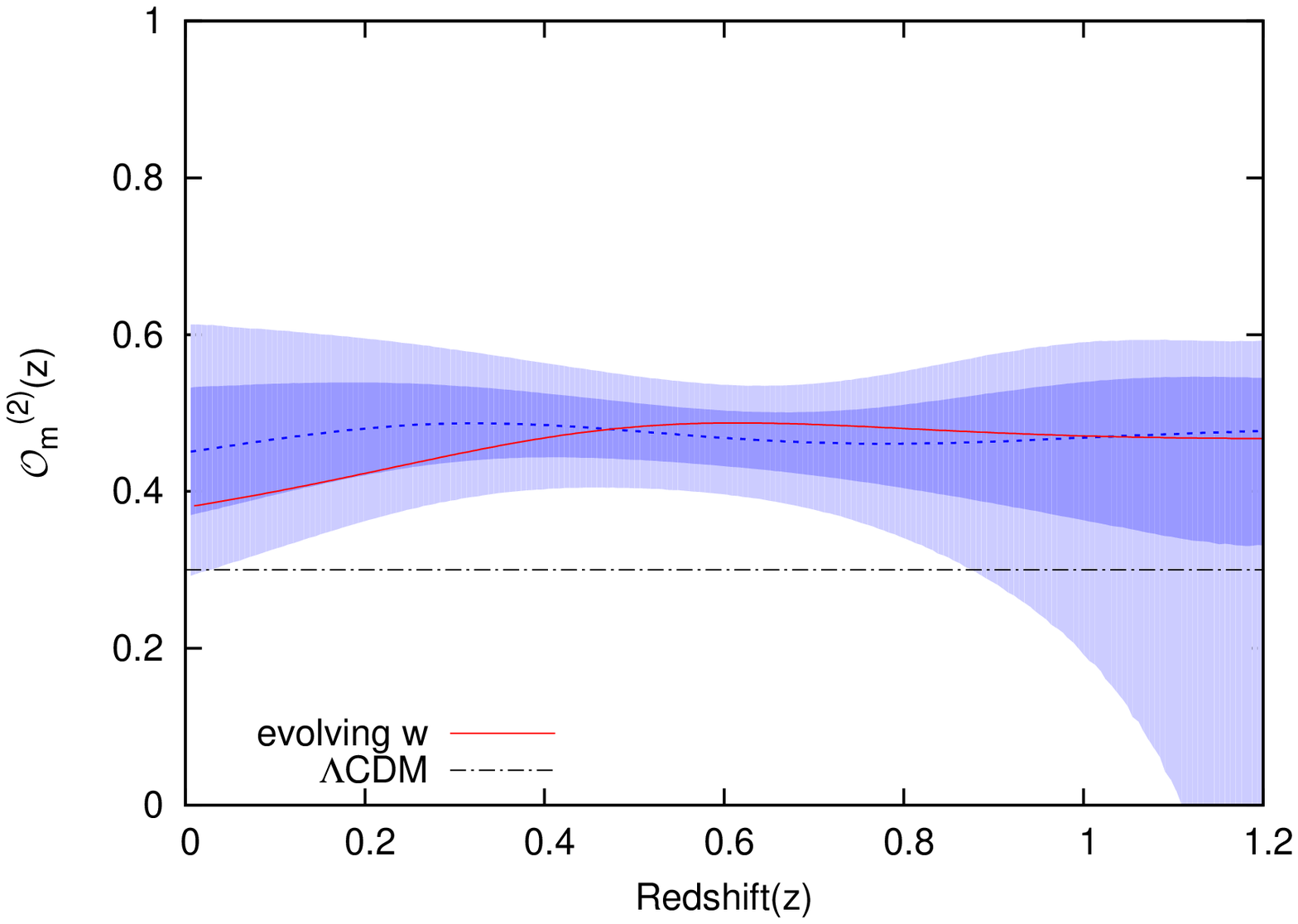}\quad
\includegraphics[width=0.3\textwidth]{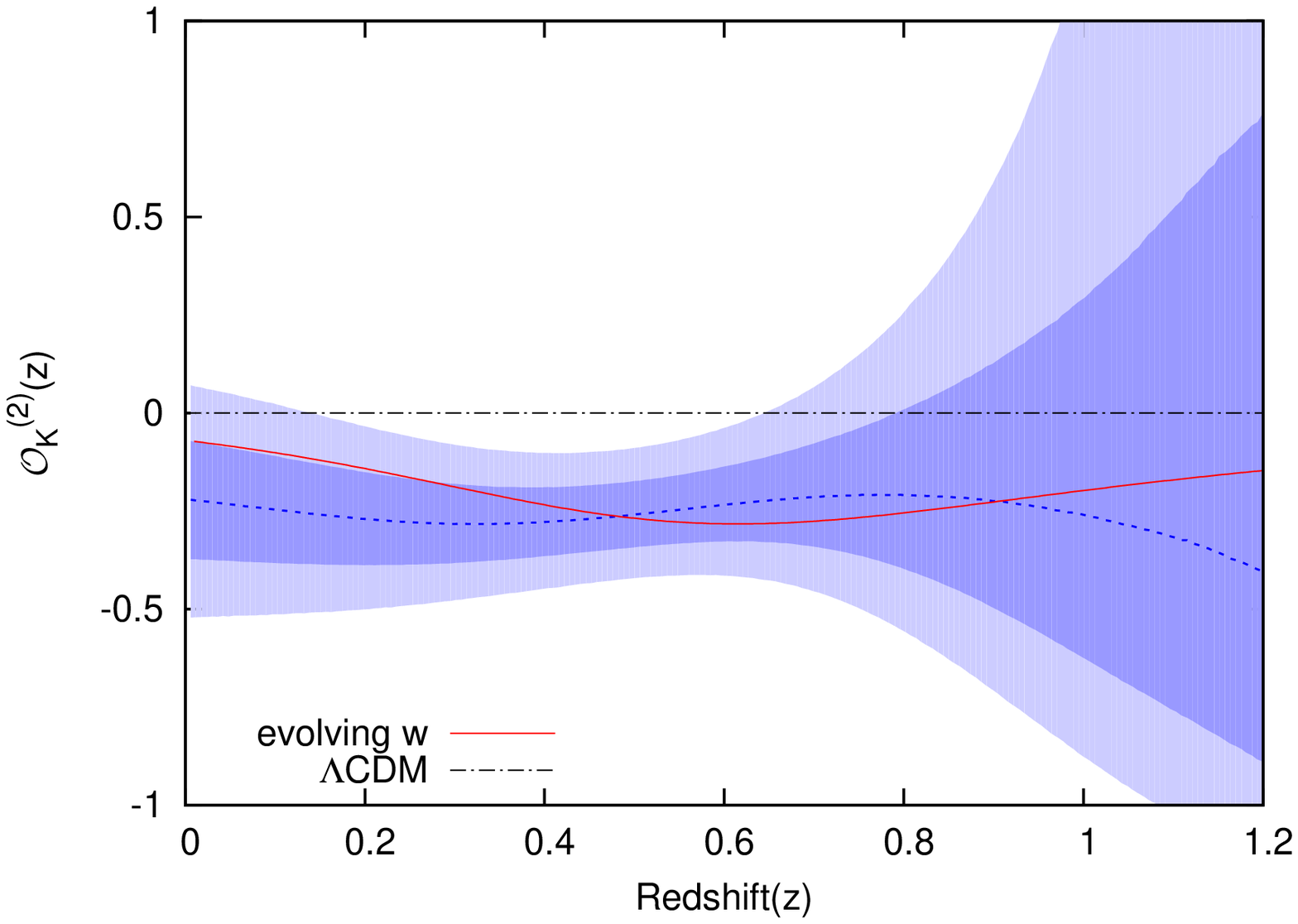}
\caption{ $\mathcal{O}_m^{(1)}$ ({\em left}), $\mathcal{O}_m^{(2)}$
  ({\em middle}) and $\mathcal{O}_K^{(2)}$ ({\em right}) reconstructed
  using simulated DES data, and assuming $\Lambda$CDM ({\em top}) and
  the evolving $w$ in \eqref{evolvingw} ({\em bottom}).  }
\label{Fig:lm11}
\end{figure*}

\begin{figure*}
\includegraphics[width=0.4\textwidth]{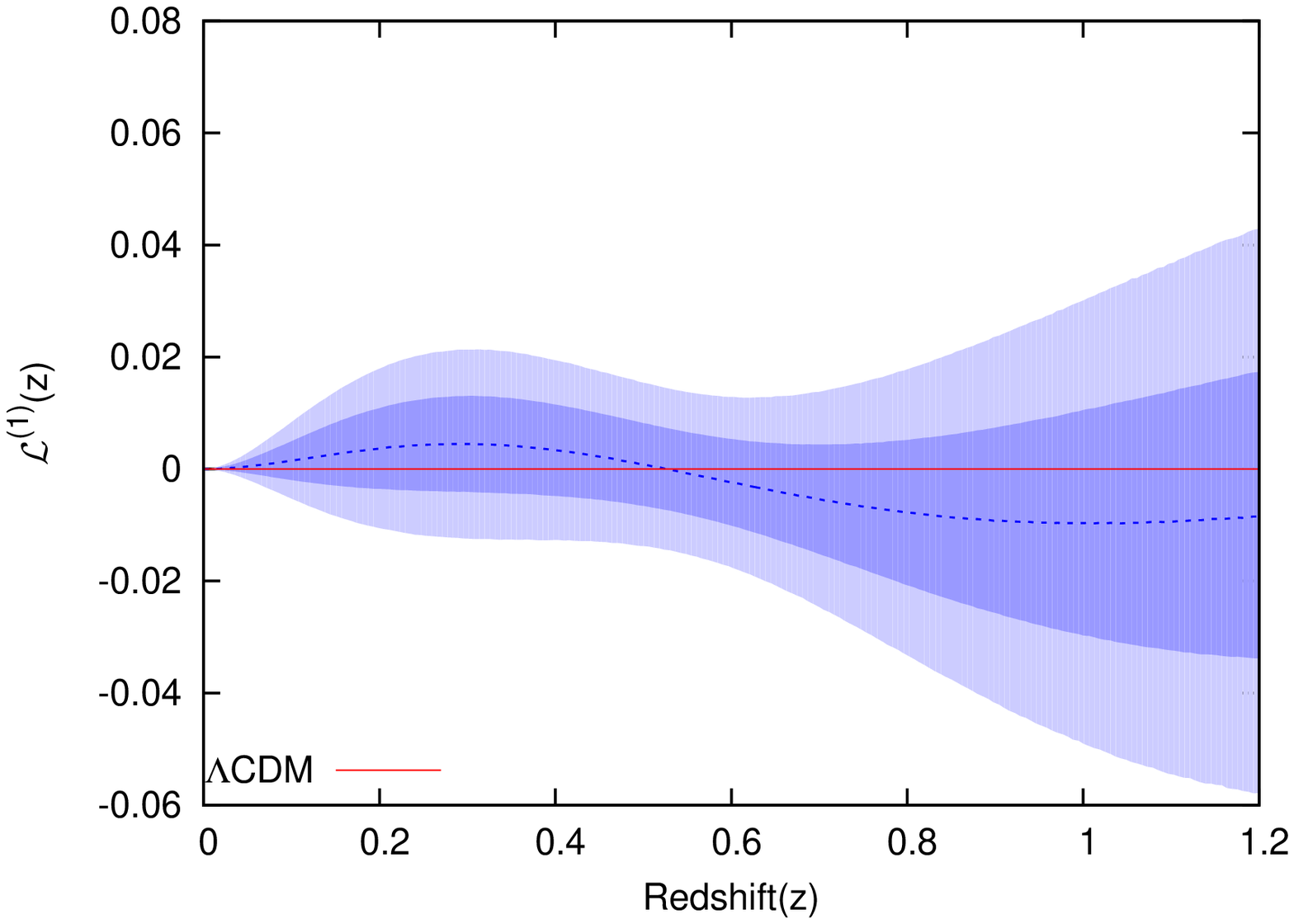}\quad
\includegraphics[width=0.4\textwidth]{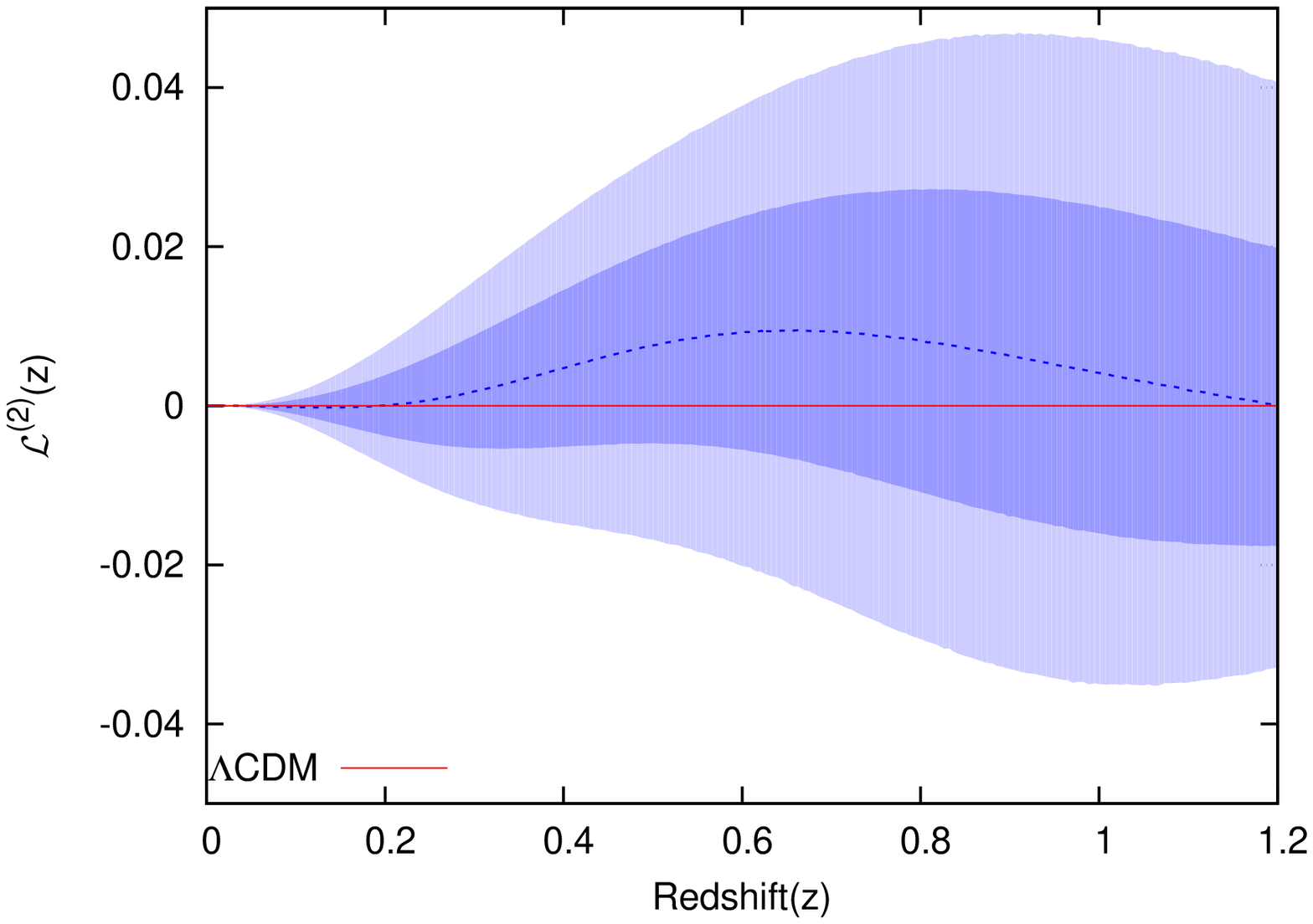}\\
\includegraphics[width=0.4\textwidth]{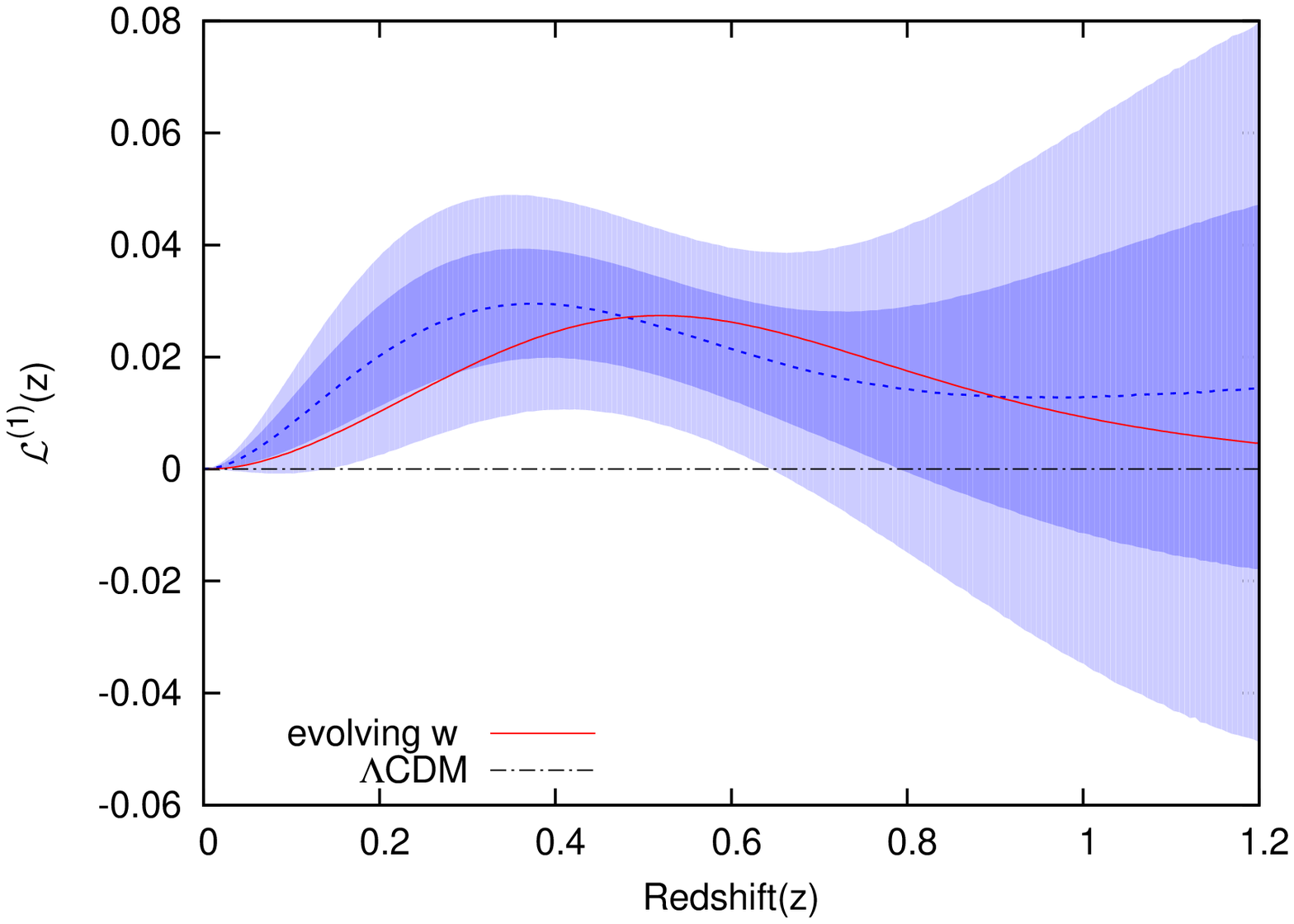}\quad
\includegraphics[width=0.4\textwidth]{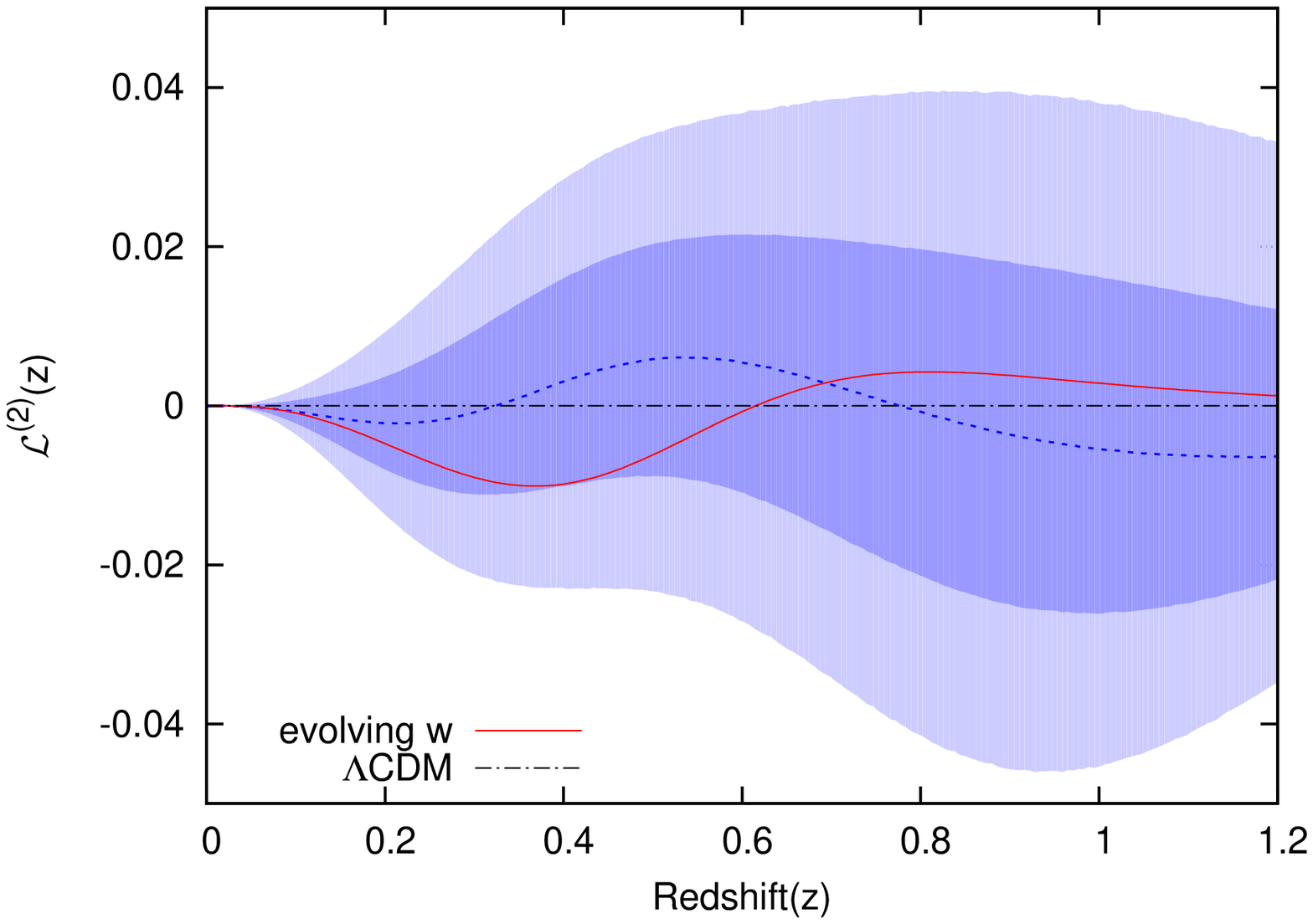}
\caption{ Reconstruction of $\mathcal{L}^{(1)}$ ({\em left}) and
  $\mathcal{L}^{(2)}$ ({\em right}) for simulated DES data, and
  assuming $\Lambda$CDM ({\em top}) and the evolving $w$ in
  \eqref{evolvingw} ({\em bottom}). Due to the degeneracy between $w$
  and $\Omega_m$, the reconstruction of $\mathcal{L}^{(2)}$ for the
  model with evolving dark energy is consistent with
  $\Lambda$CDM. However, the inferred values of $\Omega_m$ and
  $\Omega_K$ differ significantly from the input value as can be seen in
  Fig.~\ref{Fig:lm11}.}
\label{Fig:lm2_des_evlv}
\end{figure*}

For each of the two simulated data sets, we reconstruct $D(z)$ and its
derivatives and apply the null tests. Figure~\ref{Fig:lm11} shows the
constraints and uncertainties on $\mathcal{O}_m^{(1)}$,
$\mathcal{O}_m^{(2)}$ and $\mathcal{O}_K^{(2)}$ for both models, while
Figure~\ref{Fig:lm2_des_evlv} shows the results for
$\mathcal{L}^{(1)}$ and $\mathcal{L}^{(2)}$.

\section{Discussion}\label{Discussion}

We have introduced an approach to applying null tests of the
$\Lambda$CDM models (flat and curved). Using a GP technique to
reconstruct the distance-redshift relationship and its derivatives
from SNIa data sets in a model-independent fashion, we have shown that
the flat concordance model is consistent with current data, falling
within the 1$\sigma$ limits. The null tests are stronger if we assume
flatness, as expected.

For the Union 2.1 dataset, the consistency tests are in good agreement
with a constant, indicating no evidence of a deviation from a flat
$\Lambda$CDM model (see Figures~\ref{Fig:om1} and~\ref{Fig:lm1}). For
the $\mathcal{O}_m^{(1)}$ and $\mathcal{O}_m^{(2)}$ tests we find a
value for $\Omega_{{m}}\sim0.27$. $\mathcal{O}_K^{(2)}$ is consistent
with zero, as expected for flat $\Lambda$CDM. Due to the limited number
of SNIa in the Union 2.1 sample and the model-independent method we
use, the reconstructed uncertainties are significant. 

For a mock data set based on the DES supernova survey, we find that
our approach can distinguish between competing cosmological
models. Using a simulated sample drawn from a flat $\Lambda$CDM model,
the recovered distribution of $\mathcal{O}_m^{(1)}$ is constant over
the redshift range considered (Figure~\ref{Fig:lm11}), consistent with
$\mathcal{O}_m^{(1)}=\Omega_{{m}}$. For the evolving $w$ model
of~\eqref{evolvingw}, $\mathcal{O}_m^{(1)}$ deviates strongly from a
constant value, so that flat $\Lambda$CDM would be disfavoured.  This
is confirmed by the deviation of $\mathcal{L}^{(1)}$ from zero in
Figure~\ref{Fig:lm2_des_evlv}.

When spatial curvature is allowed, the constraints from the null tests
tend to be weakened, as would be expected by the degeneracy introduced
by the extra degree of freedom~\cite{Clarkson:2007bc}. For a flat
$\Lambda$CDM fiducial model, the reconstructed distribution of
$\mathcal{O}_m^{(2)}$ and $\mathcal{O}_K^{(2)}$ are consistent with
being constant and equal to $\Omega_{{m}}$ and $\Omega_K$
(Figure~\ref{Fig:lm11}), respectively, confirming that the model does
not deviate from $\Lambda$CDM, as anticipated. But the errors are
significantly larger when curvature is allowed.

For the evolving $w$ fiducial model, the reconstructions of
$\mathcal{O}_m^{(2)}$ and $\mathcal{O}_K^{(2)}$ are consistent with
constants (Figure~\ref{Fig:lm11}) -- but these constant values differ
significantly from the input values of $\Omega_{{m}}$ and $\Omega_K$,
respectively. The evolving $w$ model can erroneously be interpreted as
a $\Lambda$CDM model with a large matter density $\Omega_{{m}}$ and
negative curvature $\Omega_{{K}}$. Consequently, the reconstruction
of $\mathcal{L}^{(2)}$ (Figure~\ref{Fig:lm2_des_evlv}) is consistent
with a constant, indicating that $\Lambda$CDM is not disfavoured. In
both cases, the errors are large and the null tests are degraded.

This problem reflects the degeneracy between the density parameters
and the dark energy equation of state (see also
\cite{Clarkson:2007bc,Hlozek:2008mt,Seikel:2012uu}). The
reconstructions are formally consistent with a constant, and thus with
$\Lambda$CDM, due to their incorrectly inferred values. Additional
constraints on the value of $\Omega_{{m}}$ and $\Omega_{{K}}$ from,
for instance, BAO or CMB measurements, are needed to break this
degeneracy.

\section{Conclusions}

In this paper, we described a series of null tests that can be applied
to SNIa data to determine the consistency of observations with a
(flat) $\Lambda$CDM model -- without the need to parametrize the
equation of state of dark energy. The tests require that the distance
$D$ and the diagnostics $\mathcal{O}_m^{(1)}$, $\mathcal{O}_m^{(2)}$,
$\mathcal{O}_K^{(2)}$, $\mathcal{L}^{(1)}$ and $\mathcal{L}^{(2)}$ are
reconstructed in a model-independent way. We used GP to perform these
reconstructions.

We applied the null tests to the Union 2.1 SNIa data set. The results
were consistent with a flat $\Lambda$CDM model
(Figures~\ref{Fig:om1} and \ref{Fig:lm1}).

Using the anticipated redshift distribution for the DES supernova
survey, we produced mock data sets of 4000 SNIa, with two competing
fiducial cosmological models: flat $\Lambda$CDM and an evolving $w$
model. The reconstructed distributions of $\mathcal{O}_m^{(1)}$ for
these datasets show that the consistency tests are able to distinguish
between different cosmological models, and can correctly identify
deviations from $\Lambda$CDM, in the case when spatial flatness is
assumed. However, allowing for spatial curvature degrades the null
tests in general (although not always~-- see Fig.~\ref{Fig:om1}). The
inherent degeneracy between the equation of state of dark energy and
the density parameters ($\Omega_{{m}}, \Omega_{{K}}$) reduces our
ability to distinguish between various models. The distributions of
$\mathcal{O}_m^{(2)}$, $\mathcal{O}_K^{(2)}$ and $\mathcal{L}^{(2)}$
were consistent with a constant for the evolving $w$ model
(Figure~\ref{Fig:lm2_des_evlv}), but the inferred values of
$\Omega_{{m}}$ and $\Omega_{K}$ from the $\mathcal{O}_m^{(2)}$ and
$\mathcal{O}_K^{(2)}$ distributions were unrealistic
(Figure~\ref{Fig:lm11}). The degeneracy needs to be broken using other
data.

For future data sets which will have the power to probe $\Lambda$CDM at high precision, the null tests we have introduced will require further refinement. In particular, we need to develop a method of quantifying the significance of any possible deviation. This is left for future work.

\appendix*

\section{Redshift-dependence of the errors on the GP reconstructions}

The error of the GP reconstruction of the $n$th derivative on $D$ at
point $z^*$ is given by:
\begin{eqnarray}\label{erroreq}
&& \!\!\! \sigma\left(D^{(n)}(z^*)\right) = \left(k^{(n,n)}(z^*,z^*)\right.\\ 
&&{}- \left.K^{(n,0)}(z^*,\bm Z)\left[K(\bm
  Z,\bm Z) + C \right]^{-1}  K^{(0,n)}(\bm Z,z^*)\right)^\frac{1}{2} \nonumber
\end{eqnarray}
Here, $\bm Z$ is a vector containing the locations $z_i$ of the
data and $C$ is the covariance matrix of the data. $k$ denotes the
covariance function (here, Mat\'ern ($\nu=9/2$) as given by
eq.~\eqref{mat}) and $K$ a matrix containing covariances between the
redshift points: $[K(\bm Z,\bm Z)]_{ij} = k(z_i,z_j)$. Taking the
$n$th derivative of $k$ with respect to the first argument and the
$m$th derivative with respect to the second argument is denoted as
$k^{(n,m)}$.

Note that the first term in the equations for the errors is constant
for a given covariance function and hyperparameters. Stationary
covariance functions $k(z_i,z_j)$, such as Mat\'ern ($\nu=9/2$), only
depend on $|z_i-z_j|$, but not on $z_i$ and $z_j$
individually. Therefore, $k^{(n,n)}(z^*,z^*)$ does not depend on the
value of $z^*$.

We rewrite the equation for the errors \eqref{erroreq} as
\begin{equation}
\sigma\left(D^{(n)}(z^*)\right) = \sqrt{t_{1,n} - t_{2,n}(z^*)}
\label{erroreq2}
\end{equation}
where
\begin{eqnarray}
t_{1,n} &=& k^{(n,n)}(z^*,z^*)\nonumber\\
t_{2,n}(z^*) &=& K^{(n,0)}(z^*,\bm Z)\left[K(\bm
  Z,\bm Z) + C \right]^{-1}  K^{(0,n)}(\bm Z,z^*)\,.\nonumber
\end{eqnarray}

$t_{1,n}$ is determined by the covariance function and the
hyperparameters. It is constant in redshift and does not explicitly
depend on the data. (Note that the data are used to optimize the
hyperparameters and thus indirectly affect the value of $t_{1,n}$.)
$t_{2,n}(z^*)$ is redshift dependent and also depends on the position
and covariance matrix of the data.

\begin{figure*}
\subfloat[$D(z)$]{
\includegraphics[width=0.45\textwidth]{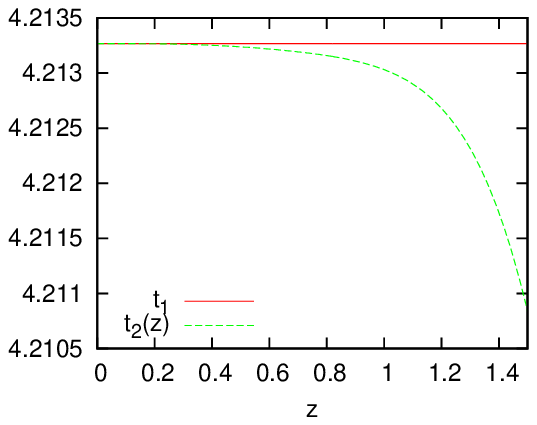}}\quad
\subfloat[$D'(z)$]{
\includegraphics[width=0.45\textwidth]{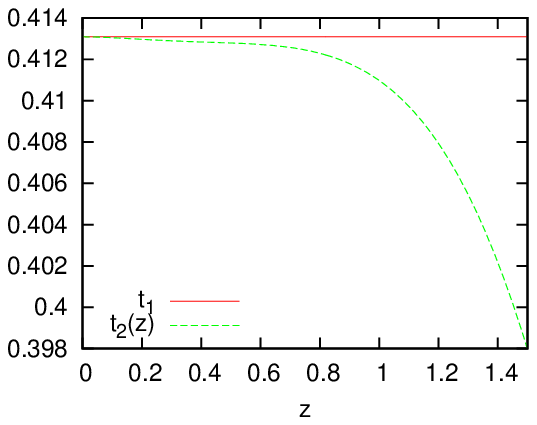}}\\
\subfloat[$D''(z)$]{
\includegraphics[width=0.45\textwidth]{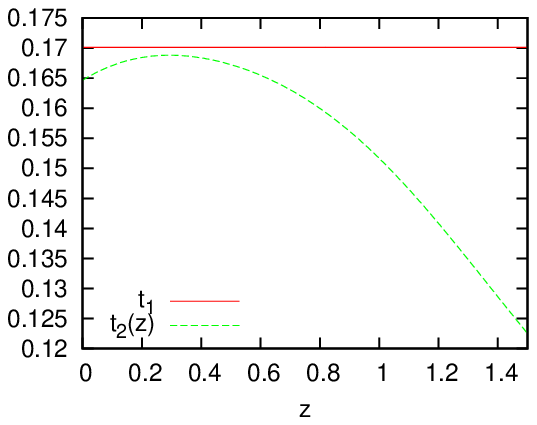}}\quad
\subfloat[$D'''(z)$]{
\includegraphics[width=0.45\textwidth]{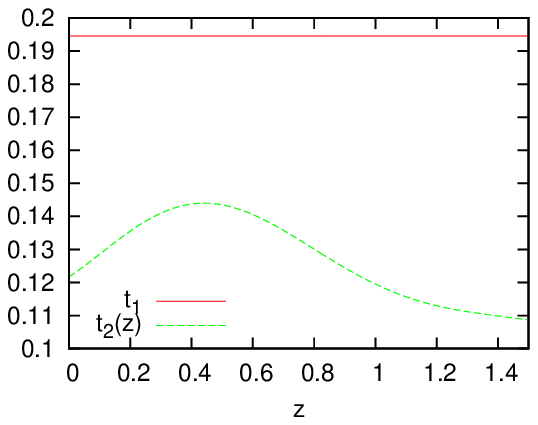}}
\caption{Terms contributing to the calculation of the errors for the
  GP reconstructions of $D(z)$, $D'(z)$, $D''(z)$ and
  $D'''(z)$. $t_1$ and $t_2(z)$ are the constant and data dependent
  terms, respectively, as given by eq.~\eqref{erroreq2}.}
\label{errorfig}
\end{figure*}

Fig.~\ref{errorfig} shows $t_1$ and $t_2(z^*)$ for the reconstructions
of $D$ and its derivatives. For $D$, $D'$ and $D''$, we observe strong
relative changes in $t_1-t_2(z^*)$ with redshift. We denote the
maximum value of this term within the considered redshift range as
$\{t_1-t_2(z^*)\}_{\text{max}}$ and the mean value as
$\{t_1-t_2(z^*)\}_{\text{mean}}$.  Then we can quantify the redshift
dependence of $t_1-t_2(z^*)$ by the relative variation
$v_r=\{t_1-t_2(z^*)\}_{\text{max}}/\{t_1-t_2(z^*)\}_{\text{mean}}$.
$v_r=1$ implies that the errors of the reconstruction are constant,
while a large value would indicate strong redshift-dependence.

We find the following results for $v_r$:
\begin{eqnarray}
v_r(D) &=&  6.9 \nonumber\\
v_r(D')  &=&  5.6 \nonumber\\
v_r(D'')  &=& 3.2 \nonumber\\
v_r(D''')  &=& 1.3 \nonumber
\end{eqnarray}
The values for $D$, $D'$ and $D''$ are much larger than 1, implying a strong
redshift-dependence of the errors.  
However, for the reconstruction of $D'''$ the relative variation is
much smaller, namely $v_r=1.3$.  The smallness of this value can be
understood by the following consideration: $t_1\gg t_2(z^*)\, \forall
z^*$ ensures that the absolute variations of $t_1-t_2(z^*)$ (which are
large compared to those for the lower derivatives of $D$) translate
into small relative variations. Therefore, the dominance of the constant
term $t_1$ is the main reason for the small value of $v_r$ and thus
the near-constancy of the errors on $D'''$.

\acknowledgments

We thank Heather Campbell for helpful discussions on SNIa data.
SY, MS and RM were supported by the South Africa Square Kilometre Array Project and  the South African National Research Foundation (NRF).  CC was supported by the NRF. RM is supported by the UK Science \& Technology Facilities Council (grants ST/H002774/1 and ST/K0090X/1). All authors were supported by a Royal Society (UK)/ National Research Foundation (SA) exchange grant.

\end{document}